%% file: main.tex
\documentclass[fleqn,10pt]{wlscirep}
\usepackage{amsmath}
\usepackage{amssymb}
\usepackage{amsfonts}
\usepackage{graphicx}
\usepackage{dsfont}
\usepackage{yfonts}
\usepackage{url}
\usepackage{subfigure}
\usepackage{comment}
\usepackage[ruled]{algorithm2e}

\usepackage[switch]{lineno}
\usepackage{multirow}
\usepackage{color}
\usepackage{makecell}
\usepackage{diagbox}
\usepackage{CJKutf8}
\usepackage{hyperref}

\usepackage{url}
 
\makeatletter
\newcommand\newcaption[1]{%
  \@namedef{ext@\@captype}{}
  \caption{#1}}
\makeatother

\title{A Machine Learning-enhanced Robust P-Phase Picker for Real-time Seismic Monitoring\footnote{Note that this paper is the English version of our work published in SCIENTIA SINICA Informationis~\cite{shen2020machinelearning}, which is suggested to be cited if needed.}}
\author[1,+]{Dazhong Shen}
\author[1,+]{Qi Zhang}
\author[1,$\ddagger$]{Tong Xu}
\author[2,$\ddagger$]{Hengshu Zhu}
\author[3]{Wenjia Zhao}
\author[1]{Zikai Yin}
\author[1]{Peilun Zhou}
\author[4]{Lihua Fang}
\author[1,$\ddagger$]{Enhong Chen}
\author[5,$\ddagger$]{Hui Xiong}

\affil[1]{University of Science and Technology of China, Hefei, Anhui, 230027, P.R. China}
\affil[2]{Baidu Inc., Beijing, 100085, P.R. China}
\affil[3]{Institute of Geology, China Earthquake Administration, Beijing, 100029, P.R. China}
\affil[4]{Institute of Geophysics, China Earthquake Administration, Beijing, 100081, China}
\affil[5]{Rutgers, the State University of New Jersey, Newark, NJ 07102, U.S.A}
\affil[+]{These authors contributed equally to this work (\{sdz,zq26\}@mail.ustc.edu.cn).}
\affil[$\ddagger$]{The corresponding authors (\{tongxu, cheneh\}@ustc.edu.cn, zhuhengshu@gmail.com, xionghui@gmail.com).}

\begin{abstract}
Identifying the arrival times of seismic P-phases plays a significant role in real-time seismic monitoring, which provides critical guidance for emergency response activities. While considerable research has been conducted on this topic, efficiently capturing the arrival times of seismic P-phases hidden within intensively distributed and noisy seismic waves, such as those generated by the aftershocks of destructive earthquakes, remains a real challenge since most common existing methods in seismology rely on laborious expert supervision. To this end, in this paper, we present a machine learning-enhanced framework based on ensemble learning strategy, \emph{EL-Picker}, for the automatic identification of seismic P-phase arrivals on continuous and massive waveforms. More specifically, \emph{EL-Picker} consists of three modules, namely, \emph{Trigger}, \emph{Classifier}, and \emph{Refiner}, and an ensemble learning strategy is exploited to integrate several machine learning classifiers. An evaluation of the aftershocks following the $MS$ 8.0 Wenchuan earthquake demonstrates that \emph{EL-Picker} can not only achieve the best identification performance but also identify 120\% more seismic P-phase arrivals as complementary data. Meanwhile, experimental results also reveal both the applicability of different machine learning models for waveforms collected from different seismic stations and the regularities of seismic P-phase arrivals that might be neglected during manual inspection. These findings clearly validate the \emph{effectiveness}, \emph{efficiency}, \emph{flexibility} and \emph{stability} of \emph{EL-Picker}. 

\end{abstract}

\begin{document}
\flushbottom
\maketitle

\input{Introduction.tex}

\input{RelatedWork.tex}

\input{Framework.tex}

\input{Data.tex}

\input{Results.tex}
\input{Discussion.tex}
\input{Conclusion.tex}

\input{bibliography_ordered.tex}
%\bibliographystyle{naturemag}
%%\bibliographystyle{named}
%\bibliography{sample}

%\section*{Author Contributions}
%D.Z.S, Q.Z, T.X and H.S.Z designed and implemented the framework of ML-Picker. D.Z.S, Q.Z, T.X, H.S.Z and W.J.Z conceived the experiment and evaluated the results. D.Z.S, Q.Z, Z.K.Y and P.L.Z processed the data and implemented the feature extraction of ML-Picker. W.J.Z and L.H.F advised on the literature review, data process and technical design of ML-Picker. T.X, H.S.Z, H.X, and E.H.C managed and advised on the project. D.Z.S, Q.Z, T.X, H.S.Z and H.X wrote the paper.

%\section*{Competing Interests and Author Information}
%Reprints and permissions information is available at www.nature.com/reprints. The authors declare no competing financial interests. Correspondence and requests for materials should be addressed to T.X (tongxu@ustc.edu.cn), H.S.Z (zhuhengshu@gmail.com), E.H.C (cheneh@ustc.edu.cn) or H.X (xionghui@gmail.com).

%\newpage
\input{supportinformation}

\end{document}

%% file: Introduction.tex
\section{Introduction}

%\subsection{Motivation}
Considerable research has been conducted on real-time seismic monitoring over the past few decades to mitigate earthquake damage. In this respect, one of the most important and relevant tasks is the identification of seismic P-phase arrivals, that is, to determine the exact arrival times of nondestructive primary earthquake waves (P-waves), which travel more quickly through the Earth’s crust than do the more destructive shear and surface waves (S-waves). If such information could be determined in real time, emergency activities, such as the activation of early warning systems and the initiation of evacuation and rescue protocols, could be conducted in a more timely manner; as a result, the damage caused by earthquakes could be largely alleviated. Although this topic has received substantial attention from seismologists, it is still very challenging to effectively identify seismic P-phases hidden within intensively distributed and noisy seismic waveforms, such as those generated by the aftershocks of destructive earthquakes, which are often uniquely characterized by waveforms with a high frequency, weak signal regularity, large magnitude range, and various natural or artificial interference factors.

In the literature, a variety of methods, such as template matching~\cite{gibbons2006detection,shelly2007non,plenkers2013low}, and higher-order statistics~\cite{saragiotis1999higher,saragiotis2002pai,kuperkoch2010automated}, have been developed for identifying seismic P-phase arrivals. 
Nevertheless, 
in seismology, guided by practical experiences, the most methods for identifying the phase arrival are developed based on the STL/LTA detector~\cite{allen1978automatic,allen1982automatic} or its variants~\cite{lomax2012automatic}, which identify earthquake events when the ratio between LTA and STA of energy function for seismic waveforms exceeds the pre-defined threshold.
However, because the implementation of these traditional approaches is generally simple, their identification performance is usually limited, due to the trade-off between false and missing alarms, 
As a result, to guarantee the correctness and avoid unnecessary panic, or decrease possible damage for public, laborious and time-consuming human supervision is still needed in real-world seismic monitoring applications~\cite{ruano2014seismic}.

Recently, various machine learning approaches been proven to achieve a competitive accuracy in pattern recognition applications and have been successfully applied for earthquake detection and seismic phase arrival picking. such as support vector machines (SVMs)~\cite{ruano2014seismic}, hidden Markov models (HMMs)~\cite{beyreuther2008continuous} and neural networks (NNs)~\cite{wang1995artificial,dai1997application,ross2018generalized,ross2018p,perol2018convolutional,zhu2018phasenet,zhang2019aftershock,mousavi2019cred,wiszniowski2014application}.
Although those machine learning-assisted methods can largely reduce labour costs, the accuracy, efficiency and stability of machine learning models cannot be fully guaranteed on continuous and massive waveforms, which restricts the application of machine learning technologies in real-time seismic P-phase arrival identification. 
More specifically, state-of-the-art studies in the literature focus on either identifying the time windows that contain earthquakes~\cite{perol2018convolutional,ross2018generalized,zhang2019aftershock}, or the capacity to determine the accurate arrival time of a seismic P-phase within a given time window containing seismic events~\cite{zhu2018phasenet,ross2018p}. 
Indeed, most of these investigations were based on a sliding window strategy, whereas few considered how to select the most appropriate time windows from real-time, continuous seismic waveform data as candidates for machine learning models. 
Unfortunately, the sliding window strategy requires extensive computational resources, preventing the deployment of sophisticated machine learning models on real-time continuous seismic monitoring systems.

\begin{figure}[!t]
\centering
\vspace{0mm}
\includegraphics[width=1.0\columnwidth]{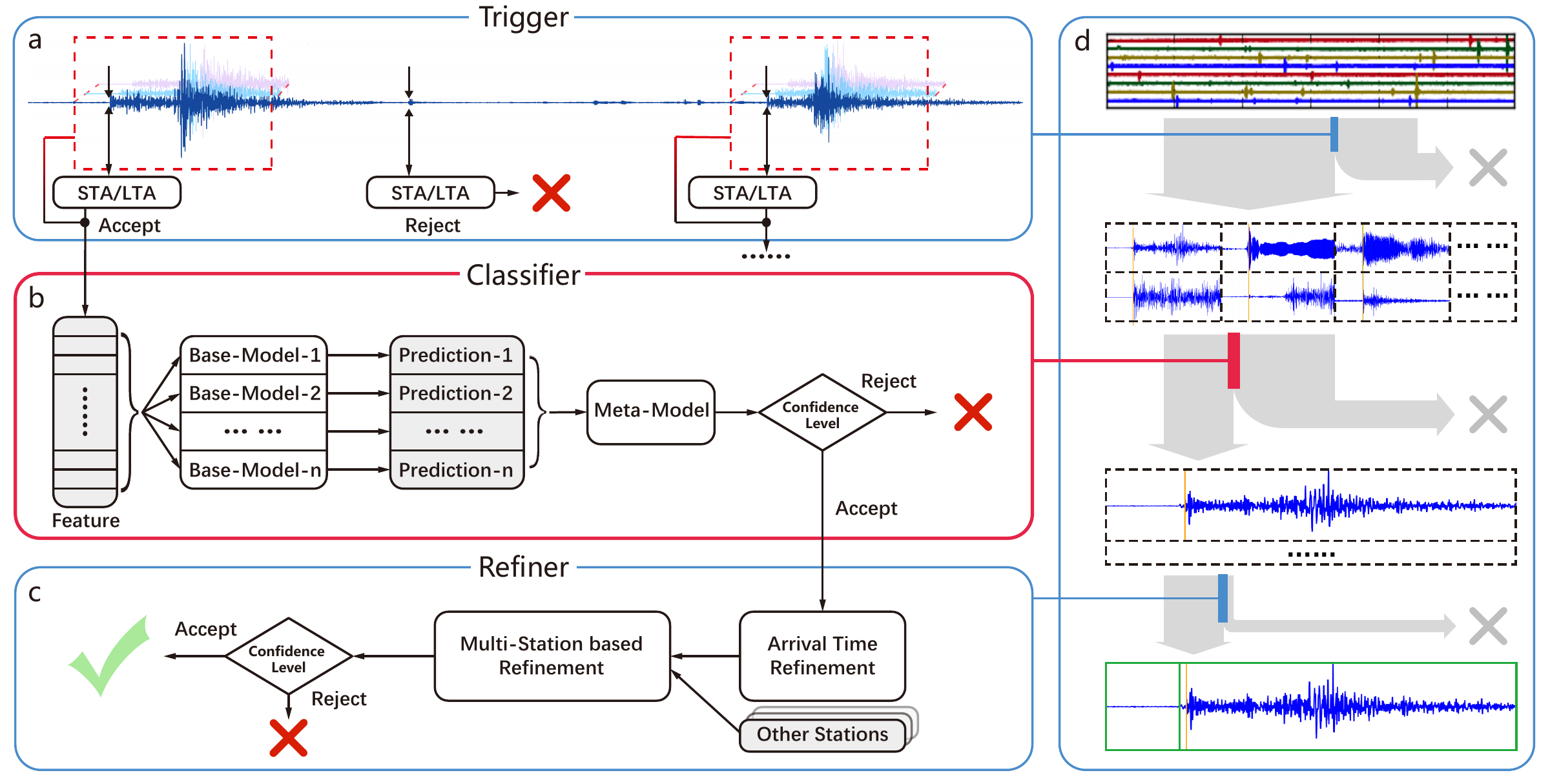}
\caption{\textbf{An overview of the \emph{EL-Picker} framework, where the machine learning-assisted parts are highlighted with red lines.} 
\emph{EL-Picker} consists of three interactive function modules, namely, \emph{Trigger}, \emph{Classifier}, and \emph{Refiner}. \textbf{(a)} With the continuous real-time waveform signals from individual stations as inputs, the \emph{Trigger} module detects potential seismic P-phase arrival candidates with traditional methods, such as STA/LTA algorithms. \textbf{(b)} The \emph{Classifier} module tries to introduce an effective discriminant model to evaluate the confidence level of each triggered P-phase candidate. With the features extracted from the waveforms near the potential arrivals, a number of distinct machine learning models are implemented to generate base model predictions. Then, a meta-model is used to integrate all these base predictions with an ensemble strategy and then output the final confidence level. \textbf{(c)} The \emph{Refiner} module identifies the most accurate arrival times from the highly confident seismic P-phase arrival candidates and rejects obvious outliers based on the Akaike information criterion (AIC) and the results from other stations. \textbf{(d)} A schematic diagram of the overall workflow in \emph{EL-Picker}, where the blue and red bars indicate the thresholds of the confidence levels in the different modules and the width of the downward arrow indicates the number of identified arrivals}
\label{fig:framework} % \label should be placed after \caption
\vspace{-2mm}
\end{figure}

To this end, in this study, we present a machine learning-enhanced framework based on ensemble learning strategy known as \emph{EL-Picker} for identifying seismic P-phase arrivals in an automatic and real-time manner. The basic idea is to first use traditional approaches, e.g., STA/LTA, which require fewer computational resources, to filter out most of the noise; then, machine learning models are used to identify seismic P-phase arrivals in the remaining time windows.
Specifically, \emph{EL-Picker} is designed in a modular manner, which consists of three components, namely, the \emph{Trigger}, \emph{Classifier}, and \emph{Refiner} modules, as shown in Figure~\ref{fig:framework}. The \emph{Trigger} module is designed to detect potential seismic P-phase arrival candidates and then filter out most noise with traditional methods, while the \emph{Classifier} module introduces an effective machine learning model to evaluate the confidence level of each triggered P-phase candidate, and the \emph{Refiner} module identifies the most accurate arrival time from the highly confident seismic P-phase candidates and rejects obvious outliers.
 In particular, an advanced Ensemble Learning strategy is designed to enhance the \emph{Classifier} module, which is well-known for the great accuracy and robustness in diverse real-world applications, and utilized generally in various rigid data mining competitions~\cite{chen2016xgboost}.
We validated the proposed framework based on the aftershocks following the $MS$ 8.0 Wenchuan earthquake by simulating a real-time seismic monitoring scenario with the continuous waveforms recorded at multiple seismic stations. The extensive experiments and discussions clearly demonstrate the \emph{effectiveness}, \emph{efficiency}, \emph{flexibility} and \emph{stability} of \emph{EL-Picker}, as well as some interesting observations. 
In particular, with the preliminary version of \emph{EL-Picker}, we won the championship in the First Season and were the runner-up in the Finals of the 2017 International Aftershock Detection Contest~\cite{fang2017seismOlympics} hosted by the China Earthquake Administration, in which 1,143 teams participated from around the world.

Finally, the contributions of this paper can be summarized as following:
\begin{itemize}[itemsep= 3 pt,topsep = 10 pt]
\item We proposed a novel three-module framework, \emph{EL-picker}, combining traditional methods and machine learning, to identify the P-phase arrival from continuous seismic waveform in an automatic and real-time manner.
\item An advanced ensemble learning strategy is well-designed to enhance the machine learning module and distinguish the seismic P-phase, which is equipped with high robustness and efficiency.
\item  The extensive experiments and discussions clearly demonstrate the \emph{effectiveness}, \emph{efficiency}, \emph{flexibility} and \emph{stability} of \emph{EL-Picker}, as well as some interesting observations. 
\end{itemize} 

%% file: RelatedWork.tex
\section{Related Work}
The related works of this paper can be grouped into two categories, namely, Traditional Methods in Seismology and Machine Learning-based Approaches.

\subsection{Traditional Methods in Seismology}
As a social good problem, the identification of seismic P-phase arrivals has attracted a wide range of research attention in the seismological communities. As a result, various methods have been proposed based on different aspects. For instance, template matching is one powerful technique, where the basic idea is that the similar waveforms may be caused by similar earthquake mechanism. Gibbons, et al.~\cite{gibbons2006detection} are among the first to apply the template matching in the field of detection seismology. Then, a large amount of attempts, such as autocorrelation~\cite{brown2008autocorrelation,aguiar2014pagerank}, similarity searching~\cite{yoon2015earthquake} and etc.~\cite{shelly2007non,plenkers2013low}, have been proven as sensitive and discriminative solutions for finding similar earthquakes. However, those approaches usually suffer expensive computational burden and cannot be implemented on real-time seismic phase picking. 
In addition, several statistical approaches have also been used to process the seismic waveform, such as  higher order statistical functions (HOS). 
Saragiotis et al. ~\cite{saragiotis1999higher,saragiotis2002pai}  are one pioneer to utilize HOS functions (HOS) to P-phase arrival time determination, introducing the skewness and the kurtosis functions to phase picking. Then, Kuperkoch et al.~\cite{kuperkoch2010automated} developed Saragiotis’ method and designed a quality-weighting scheme for picks.  
However, guided by practical experience, the most comment methods applied in real-time seismic monitoring are designed following by the STA/LTA detector, as well as its variations, e.g., Allen Picker~\cite{allen1982automatic}, BK87~\cite{baer1987automatic} and etc.~\cite{earle1994characterization}. Among them, the FilterPicker algorithm developed in ~\cite{lomax2012automatic} has relatively robust and efficient performance on seismic phase picking, where the energy function for seismic waveforms has been deliberately designed  and is more sensitive for seismic events.

\subsection{Machine Learning-based Approaches}
In the recent decades, substantial researches have involved the machine learning approaches into many fields of seismology successfully, such as earthquake location~\cite{wu2019research}, earthquake magnitude~\cite{asim2017earthquake}, earthquake classification~\cite{riggelsen2014machine}, and intensity estimation~\cite{zhu2020rapid}. 
In particular, for earthquake detection and seismic phase arrival picking, a variety of machine learning approaches have been discussed and proven to achieve a competitive accuracy. Specifically, 
Ruano, et al.~\cite{ruano2014seismic} designed a seismic detection system with Suport Vector Machine (SVM) and Multi-Layer Perceptrons (MLPs).
Moritz, et al.~\cite{beyreuther2008continuous} applied hidden Markov models (HMMs) classifier to the detection and distance dependent classification of small to medium sized earthquakes.
Among them, Neural network is one powerful and popular approach.
Since the idea proposed by Wang, et al.\cite{wang1995artificial} to apply fully-connected neural networks to seismic event detection, neural network has been widely studied in this filed~\cite{dai1997application,ross2018generalized,ross2018p} due to the nature of flexibility and generalization~\cite{gentili2006automatic}.
Besides the classic neural network, conventional neural network have also been involved to improve the detection performance recently.
Perol, et al. ~\cite{perol2018convolutional} designed one CNN structure for earthquake detection from time window candidates.
Zhu, et al.~\cite{zhu2018phasenet} applied CNN to generate the probability distribution of phase arrivals, which plays the similar role as the energy function in traditional STA/LTA methods.
And, the CNN structure in Zhang, et al.~\cite{zhang2019aftershock} can capture both long-term scale and short-term scale seismic features.
In addition, in spired by the analogy between sound signal and seismic signal, researchers also attempted to utilize Long Short-Term Memory Networks (LSTM), which has been utilized widely in speech recognition~\cite{graves2013hybrid}.
For instance, Mousavi, et al.~\cite{mousavi2019cred} and Wiszniowski, et al.~\cite{wiszniowski2014application} introduced earthquake detection systems enhanced by LSTM, and demonstrated their advantages for microearthquakes detection.
However, 
those approaches usually focus on either identifying the time windows that contain seismic phase from time window candidates~\cite{perol2018convolutional,ross2018generalized,zhang2019aftershock}, or determining the accurate arrival time of a seismic P-phase within a given time window containing seismic events~\cite{zhu2018phasenet,ross2018p}.
Moreover, compared with traditional STA/LTA methods, machine learning approaches require extensive computation, therefore it is not practical to identify every time window, or every signal point, without selecting potential candidates, which preventing the deployment of sophisticated machine learning models on real-time continuous seismic monitoring systems.

%% file: Framework.tex
\section{EL-Picker Framework}\label{sec:framework}
\begin{figure}[!tphb]
	\centering
    %\vspace{-3mm}
    \includegraphics[width=0.85\columnwidth ]{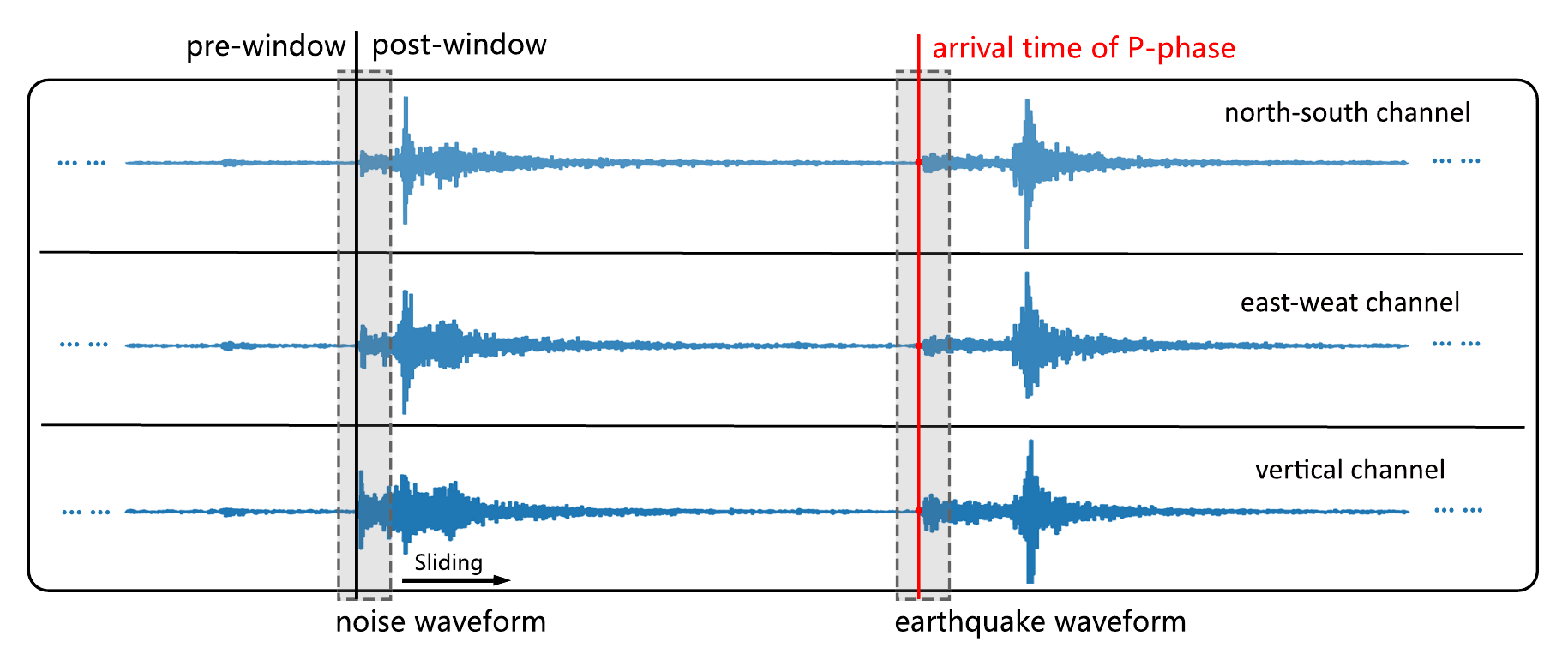}
    \caption{\textbf{An example for the real-world seismic waveforms on three dimensions} The red solid line indicates the P-phase arrival for one earthquake. As a comparison, the black solid line indicates the beginning of one noise  event.The grey dashed blocks represents the short time windows around current signal point, which can be further split into the pre-window and post-window.}
    \label{fig:defexample}
\vspace{-2mm}
\end{figure}

Generally, seismic waveforms are recorded on three channels corresponding to three spatial dimensions, i.e., Z for the vertical channel, N for the north-south channel, and E for the east-west channel.  Figure~\ref{fig:defexample} shows one period of the real-world seismic waveforms on those three dimensions as an example. (more examples are shown in Figure~\ref{fig:examplefordata} in~\ref{SI:sec:detailinformation}) We can observe a phase of waveform vibration spurred by one seismic event. And due to the primary waves (P-waves) always reach the earth surface and cause several vibrations in three channels first, seismic researchers usually denote the beginning point of those vibrations (tagged by red solid line) as the arrival time of P-phase, which indicated the occurrence of one earthquake event. In addition, there always exist noise waveforms in seismic waveforms which are extremely difficult to be distinguished from seismic event waveforms. Intuitively, the identification of one P-phase arrival requests rigorous analysis for the waveform vibrations in the short time window around itself. Along this line, our target of P-phase picking can be defined as: Given the record of continuous real-world seismic waveforms on three channels, our goal is to automatically pick up the exact arrival time of P-phase for each involved seismic event from various noise signals, based on the short time window around each signal point.

As mentioned above, three components are involved in our proposed \emph{EL-Picker} framework, namely, the \emph{Trigger}, \emph{Classifier}, and \emph{Refiner} modules, as shown in Figure~\ref{fig:framework}. In the following, we will separately introduce their technical details.

\subsection{Trigger Module}
To guarantee a suitable identification efficiency, in the \emph{Trigger} module, we propose to leverage traditional methods by implementing STA/LTA algorithm or its variants on energy functions for the real-time waveform signals from individual stations. For every change on energy functions that is beyond a pre-defined threshold, a potential seismic P-phase arrival candidate will be selected. Note that to capture as many seismic P-phase arrivals as possible, the threshold should be relatively small.

To be specific, in this paper, the \emph{Trigger} module was implemented by \emph{FilterPicker}~\cite{lomax2012automatic}, a broadband phase picking algorithm that is loosely based on the STA/LTA algorithm. The first step in \emph{FilterPicker} is to perform multi-bandpass filtering on the signal. In the validations, we utilized three bands (2.5-5 Hz, 5-10 Hz, and 10-20 Hz) of waveform signals on channel Z. Then, an energy function, which includes several parameters, namely, the triggering threshold $S_1$, the time width $T_{up}$ and the average threshold $S_2$, was defined to monitor each time step to check for triggers or picks. At a certain moment $t$, if the function value of $t$ exceeds $S_1$ and the average function value during the time window $T_{up}$ after t exceeds $S_2$, the moment $t$ will be recorded as a potential P-phase arrival. In particular, the thresholds were set as $S_1=6$, $S_2=2$ and $T_{up}=0.3$ (seconds), which are smaller than that in the literature~\cite{lomax2012automatic}.

 \subsection{Classifier Module}
Due to relatively small threshold in the \emph{Trigger} module, the number of false alarms will be also relatively high. Therefore, we must further filter out positively labelled false candidates and identify seismic P-phase arrivals with a high confidence and reliability. Therefore, once a potential arrival has been triggered, a time window of the waveforms near the arrival time will be conveyed to the \emph{Classifier} module, which will effectively distinguish seismic P-phases from other signals (e.g., noises) with the features extracted within. In particular, to guarantee robustness, the \emph{Classifier} module is also designed in a modular manner with ensemble learning~\cite{rokach2010ensemble}; that is, a number of distinct base models are implemented and integrated as a meta-model, which is also known as the stacking strategy~\cite{wolpert1992stacked}. Here, a greater number of seismic P-phase arrivals would obtain a higher score from the meta-model as the confidence level. The fundamental idea of ensemble learning is to construct a predictive model by integrating multiple machine learning models, thereby improving the prediction performance~\cite{rokach2010ensemble}. In the \emph{Classifier} module, each base model (e.g., an SVM~\cite{cristianini2000introduction}) can be regarded as a theory for predicting the probability that one candidate is the true seismic P-phase arrival. However, in most cases, each individual theory contains certain biases that cause prediction errors; for example, SVMs with a linear kernel cannot process nonlinear feature components~\cite{cristianini2000introduction}. Therefore, the stacking strategy used in \emph{EL-Picker} tries to minimize this error by reducing these biases with respect to the provided learning set. The central concept is that we can do better than simply list many ``theories'', which are consistent with the learning set, by constructing an optimal ``theorist'' that combines all those ``theories''~\cite{wolpert1992stacked}. Specifically, in \emph{EL-Picker}, for each candidate sample, the meta-model integrates the predictions from all base models by using a linear model (logistic regression~\cite{cox1958regression} was employed in our experiments), which can weigh each base model with linear coefficients. Accordingly, the base models can be improved through independent adjustments, and the \emph{Classifier} module itself can boast a better performance through the selection of preferred base models.

In our implementation, 9 commonly used classifier models, namely, SVM-linear, SVM-poly~\cite{ruano2014seismic}, Tree-gini, Tree-entropy~\cite{breiman2017classification}, K-nearest Neighbors (KNN), RandomForest~\cite{ho1995random}, Adaboost~\cite{freund1997decision}, Logistic Regression~\cite{cox1958regression}, and Gaussian Naive Bayes~\cite{russell2016artificial}, were intuitively selected as the base models.
The complete training and validation stages were designed as follows:
\begin{itemize}
\item All the training data were divided into 5 parts. For each base model, we applied a 5-fold cross-validation~\cite{kohavi1995study} procedure on training data.
To be specific, we employed the data in each part as the test data, while the data in the other four parts were utilized as the training data in five rounds of testing.  Then, each sample in each part could obtain a score from each base model as the judgement whether one candidate is the true seismic P-phase arrival.
\item As each sample was assigned 9 judgements (i.e., by the 9 machine learning models mentioned above), logistic regression was performed to achieve the weights for each base model based on the judgements and corresponding labels.
\item We trained the \emph{best solution} of each base model on all the training data and then combined those solutions based on the weights achieved in the previous step.
\item In the validation stage, for each time window of waveforms, the \emph{Classifier} module outputs its confidence level score, and the windows with higher scores than the pre-defined threshold (set to 0.5) are regarded as potential seismic P-phase arrivals.
\end{itemize}

\vspace{-1mm}
\subsection{Refiner Module}
Although the above mentioned modules can identify seismic P-phase arrivals with a high confidence level, some avenues remain with which to improve the results. For example, STA/LTA algorithms are usually not sufficiently sensitive to identify exact arrival times. Moreover, while a seismic event could be intuitively monitored by utilizing multiple adjacent stations, other modules take advantage of waveform signals from only individual stations, and thus, some misleading results might be obtained. Therefore, in the final step, we introduce the \emph{Refiner} module to filter out the outliers and select the most accurate seismic P-phase arrivals as outputs.
Specifically, we first propose to use the Akaike information criterion (AIC)~\cite{akaike1974new}, which has been long utilized for picking and refining seismic phase~\cite{sleeman1999robust,leonard1999multi,leonard2000comparison}, to refine the arrival time of each candidate. Then, for each identified seismic P-phase arrival, all other seismic stations that have also monitored the P-phase arrival are regarded as indicators with which to measure the multi-station-based confidence level. To be specific, those two parts constituting this module were performed as follows:
\begin{itemize}
\item The Akaike information criterion (AIC)~\cite{akaike1974new} was applied to refine the arrival time of each candidate within a short window of $\pm 1$ second near the original time point of each P-phase arrival.
\item For any pair of arrivals on different stations, we checked whether their time difference was shorter than the maximal P-wave propagation time approximated as $D/v_p$. Here, $v_p$ was set as $5.5km/s$, which is less than the average P-wave velocity in Sichuan to tolerate some error; and $D$ represents the distance between two seismic stations corresponding to the pair of arrivals. All P-phase arrivals that were recorded by only one monitoring station were removed.
\end{itemize}

%% file: Data.tex
\section{Data Description and Feature Extraction}\label{sec:data}
In this paper, two data sets are involved, i.e., the aftershocks of the $MS$ 8.0 Wenchuan Earthquake (Sichuan, China, May 12, 2008) and $MW$ 6.0 South Napa Earthquake (California, USA, August 24, 2014).
These data sets were provided by the Data Management Centre of the China National Seismic Network at the Institute of Geophysics, China Earthquake Administration (CEA). More specifically, all the aftershocks were recorded at 100 Hz on channels Z, E,and N . To ensure the quality of the data, all the records were automatically checked to remove some erroneous fragments, which usually contain mechanical noise or even no signal at all. Accordingly, a total of 12,696 P-phase arrivals in the Wenchuan data set were labelled over a course of 4 weeks; 3,447, 3,442, 2,997 and 2,810 P-phase arrivals were picked in each successive week.  At the same time, 680 P-phase arrivals were recorded in Napa data set within 4 days.
Subsequently, to train and evaluate the \emph{Classifier} module, we selected five times more negative samples from potential candidates located far away from the labelled arrivals (more than 0.4 seconds) that were previously picked by the \emph{Trigger} module. These negative samples included all types of waveform signals, such as noise, inaccurate seismic P-phase arrivals, and seismic S-phase arrivals, which mean the arrival times of surface waves (S-waves) of earthquakes.

\begin{table}[!b]
\centering
  \renewcommand\arraystretch{1.5}
  \caption{\textbf{Detail Computation Procedures for Classification Features.} The interval in the column, titled by ``time window'', represents the selected short time windows for computation of specific features, where the time point of arrival is tagged by 0,  the negative number means the length of pre-windows and the positive number means the  length of post-windows.$AN>5$ is the lengths of post-windows in the cut time windows.}
    \resizebox{1.0\linewidth}{!}{
    \begin{tabular}{|c|c|c|c|c|c|}
    \hline
     &\makecell{Time window \\/second} & \makecell{Bandwidth \\/HZ}& Channel & \makecell{Computation process \\ for each time window $x$ \\ with each bandwidth \\ in each channel}& Number of features\\ \hline \hline	
    Amplitude Fluctuation & \makecell{(-5, 0), (0, AN),\\ (-1, 0), (0, -1)\\ (5(i-1), 5i)\\$1\leq i \leq int(AN/5)$)} &\makecell{2-10\\10-20}&\makecell{E\\ N\\ Z}& \makecell{mean(x)\\ var(x)} & \makecell{48+\\12$\times int(AN/5)$}\\ \hline

    \multirow{2}{*}{\makecell{Maximal Amplitude}}&\multirow{2}{*}{\makecell{(2, AN)}}&\multirow{2}{*}{\makecell{2-10\\10-20}} &\makecell{E\\ N}&  \makecell{idx=x.index(Max(x))\\mean(x[idx-1, idx+1])\\var(x[idx-1, idx+1])}& 12\\ \cline{4-6}
    &&&Z & \makecell{idx=x.index(Max(x))}& 2\\ \hline
    Spectral Waterfall &\makecell{(-0.2, 0), (0, 0.2)\\(-0.4, 0), (0, 0.4)\\(-0.6, 0), (0, 0.6)\\(-0.8, 0), (0, 0.8)\\(-1.0, 0), (0, 1.0)} &\makecell{0.5-0.833\\0.833-1.389\\1.389-2.314\\2.314-3.858\\3.858-6.430\\6.430-10.717\\10.717-17.816\\17.816-29.768\\20.768-49.615}&\makecell{E\\ N\\ Z}& \makecell{mean(x) \\ var(x)}& 540\\ \hline

    \multirow{2}{*}{\makecell{Other Features}} &\multirow{2}{*}{\makecell{(-5, 5)}}&\multirow{2}{*}{\makecell{1.389-2.314\\2.314-3.858\\3.858-6.430\\6.430-10.717\\10.717-17.816}}&\makecell{E\\ N\\ Z}& \makecell{RMS Amplitude Ratio(x[0, 5], x)\\Mean Difference(x[0, 5], x) \\ Envelope Slope(x)}& 60\\ \cline{4-6}
    &&&\multicolumn{2}{c|}{\makecell{Polarization Slope($x[-5,5]$ in channel E, \\ N, Z)\\ for each time window \\ with each bandwidth}}& 5\\ \hline
    \end{tabular}}
\label{tab:FeatureComputation}
\end{table}

Here, we summarize the features utilized in the \emph{Classifier} module. Specifically, for each pick within the seismic sequence to be judged, a short time window is cut to extract features for subsequent machine learning-assisted classification. Additionally, the length of the cut time window is adjustable according to the desired accuracy. 
Here, we set the length of pre-windows before the potential arrivals as 5 seconds. Then, for different lengths of post-windows after the potential arrivals from 5 to 20 seconds, we extracted \textbf{679-715} different features to characterize this pick. As Table~\ref{tab:FeatureComputation} shows, these features can be roughly divided into 4 categories of features as follows: 

\begin{itemize}
\item \textbf{Amplitude Fluctuation}. Considering the sharp fluctuations in the amplitudes of the P-phase arrivals, we selected \textbf{5-8} short time windows (according to post-window length) around the arrivals, and then filtered the waveforms on all the \textbf{3} channels with \textbf{2} bandwidths (2-10Hz and 10-20Hz) to compute \textbf{2} values, namely the \textbf{mean value} and the \textbf{variance} of amplitude. In summary, we contained \textbf{60-96} features for amplitude fluctuation.

\item \textbf{Maximal Amplitude}. Considering that the S-phase usually follows the P-phase with a maximal amplitude, especially on the horizontal channels. Thus, we summarized the maximal amplitude on all \textbf{3} channels, resulting in 3 features. Additionally, we computed the \textbf{mean value} and \textbf{variance} of the amplitude within a window of $\pm 1$ second around the maximal amplitude, on both the N (north-south) and the E (east-west) channels; consequently, we obtained additional 4 features. Moreover, the features were extracted with \textbf{2} bandwidths (2-10Hz and 10-20Hz), which results in a total of \textbf{14} features in total.

\item \textbf{Spectral Waterfall}. Considering the importance of the spectral waterfall of waveforms around the P-phase arrival, we selected a short window of $\pm 1$ second around each estimated arrival, and then divided it into \textbf{10} segments. For each segment, we filtered the waveform with \textbf{9} adjacent bandwidths lower than 50Hz to compute \textbf{2} values, namely the \textbf{mean value} and the \textbf{variance} of the amplitude. The features were extracted on all the \textbf{3} channels, which results in  total of \textbf{540} features in total.

\item \textbf{Other Feature}. Some other features were extracted to enhance the classification, such as amplitude ratio, mean difference, the polarization slope and envelope slope of the arrivals (the detail computation process can be found in Table~\ref{tab:functionsforfeature} in~\ref{SI:sec:detailinformation}).  In total, \textbf{65} additional features were extracted. 
\end{itemize}

%% file: Results.tex
\begin{table}[!b]
\center
\caption{\textbf{The performance of the \emph{Classifier} module and the base models.} The table shows the average performances of the overall \emph{classifier} module ,baselines and each base model on the 4-fold cross validation procedures when the length of post-window is set as 20 seconds. ``(N)''indicates the validation on Napa data set, while other validations are conducted on Wenchuan data set.}
\setlength{\tabcolsep}{1.8mm}{
\begin{tabular}{cccc|cccc}\hline\hline
        Methods     & Precision & Recall & F-value &    Methods   & Precision & Recall & F-value \\\hline
LSTM         & 0.7918     & 0.7494  & 0.7682 & SVM-linear         & 0.8832    & 0.8932 & 0.8878  \\
 Inception    & 0.8122    & 0.8144 & 0.8124  & LogisticRegression & 0.8822    & 0.8939 & 0.8876  \\
ConvNetQuake & 0.8808    & 0.8346 & 0.8563  & Adaboost           & 0.8861    & 0.8869 & 0.8860  \\
Classifier   & \textbf{0.9027}    & \textbf{0.8869} &\textbf{0.8941}  & RandomForest       & 0.8816    & 0.8828 & 0.8814  \\\cline{1-4}\cline{1-4}
\multicolumn{4}{c|}{{Testing on Napa data set}}   & SVM-poly           & 0.8703    & 0.8101 & 0.8384  \\ \cline{1-4}
LSTM(N)       & 0.1511     & 0.4328  & 0.2240    & Tree-entropy       & 0.8336    & 0.8319 & 0.8321  \\
Inception(N)    & 0.9708    & 0.1976 & 0.3284   & Tree-gini          & 0.8237    & 0.8266 & 0.8246  \\
ConvNetQuake(N) & \textbf{0.9848}   & 0.1940 & 0.3242   & KNN                & 0.7839    & 0.7831 & 0.7824  \\
Classifier(N)   & 0.8059    & \textbf{0.9226} & \textbf{0.8603}& NaiveBayes & 0.6956    & 0.8876 & 0.7789 \\ \hline\hline
%\multicolumn{8}{l}{\footnotesize{``(N)'' indicates the validation on Napa data set, while other validations are conducted on Wenchuan data set.}}\\
\end{tabular}}
\label{tab:performance1}
\end{table}

\section{Results}\label{sec:results}
To comprehensively and intuitively evaluate the performance of \emph{EL-Picker}, we applied the typical 4-fold cross-validation procedure to the Wenchuan data set.
More specifically, we employed the data in each week as the test data, while the data in the other three weeks were utilized as the training data in four rounds of testing.
Moreover, we implemented an additional experiment on the Napa data set with all the four weeks in Wenchuan data set as training samples, to validate the robustness of \emph{EL-Picker}.
In addition, we applied \emph{Precision}, \emph{Recall} and \emph{F-score} as the metrics. 

In the first set of validations, we attempted to evaluate the individual performance of the \emph{Classifier} module in \emph{EL-Picker}. To that end, we treated the aftershocks that were manually captured by experts as \emph{positive} samples. Correspondingly, five times more \emph{negative} samples were manually selected from the seismic sequences. The average performances of the \emph{Classifier} module are summarized in the left half of Table~\ref{tab:performance1}, where the length of post-window is set as 20 seconds.
To produce some baselines with the same experimental settings for comparison, we implemented three machine learning-assisted approaches: 
LSTM Net~\cite{hochreiter1997long}, which is one of state-of-the-art neural network for classification of time series signals;
Inception Net~\cite{xie2017aggregated}, which is one of state-of-the-art convolutional neural network methods for classification;
ConvNetQuake~\cite{perol2018convolutional}, which is a highly scalable and effective convolutional NN (CNN) for earthquake detection applications.

The results clearly reveal that the \emph{Classifier} module can achieve a competitive performance in distinguishing between the true and false samples of P-phase arrivals within the cut time windows and consistently outperforms other baselines, especially,in term of F-score, which comprehensively considers Precision and Recall through their harmonic average. 
More interestingly, the \emph{Classifier} module also achieves relatively robust performance when testing on the Napa data set, while other baselines suffer a large performance degradation, especially in term of Recall. It clearly demonstrates the effectiveness of the \emph{Classifier} module.
Furthermore, we can realize that the performance on the Napa data set seems worse than the others in Wenchuan data set in term of all approaches. It is reasonable when considering different characteristics of earthquakes caused by the different locations and different geological environment between Wenchuan and Napa.
Indeed, with ensemble learning, our \emph{Classifier} module can also achieve better performance than other  machine learning models that were selected as base models. As right half of Table~\ref{tab:performance1} shows, the overall \emph{Classifier} module outperforms each base model under Precision and F-score criteria, although they all achieve competitive performance with our deliberately extracted features.

\begin{figure}[!t]
\centering
\begin{minipage}[c]{0.48\columnwidth}
\centering
\includegraphics[width=1.0\columnwidth ]{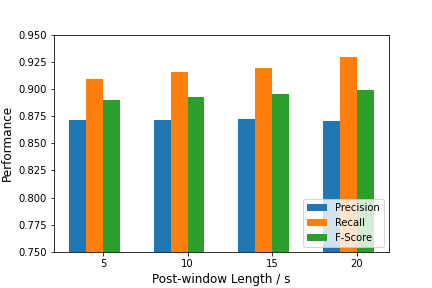}
\centerline{(b)}
\end{minipage}
\hspace{0.02\textwidth}
\begin{minipage}[c]{0.48\columnwidth}
\centering
\includegraphics[width=1.0\columnwidth ]{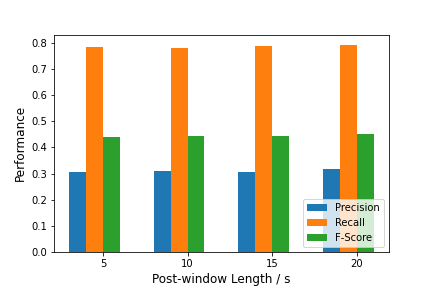}
\centerline{(b)}
\end{minipage}
\caption{\textbf{The sensitivity to the time window length within the \emph{Classifier} module (a) and the \emph{EL-Picker} framework (b).}}
\label{fig:timewindows}
\vspace{-2mm}
\end{figure}

%\begin{table}[!t]
%\center
%\caption{\textbf{The performance of the \emph{Classifier} module compared with based models.} The table shows the average performances of the overall \emph{classifier} module and each base model on the 4-fold cross validation procedures when the length of post-window is set as 20 seconds.
%}
%\begin{tabular}{cccc}
%\hline
%&Precision&Recall&F-Score\\\hline
%\emph{Classifier} Module&\textbf{0.9027} &0.8869 &\textbf{0.8941}  \\ \hline
%SVM-linear&0.8832 &0.8932 &0.8878  \\
%LogisticRegression&0.8822 &\textbf{0.8939} &0.8876  \\
%Adaboost&0.8861 &0.8869 &0.8860  \\
%RandomForest&0.8816 &0.8828 &0.8814  \\
%SVM-poly&0.8703 &0.8101 &0.8384  \\
%Tree-entropy&0.8336 &0.8319 &0.8321  \\ 
%Tree-gini&0.8237 &0.8266 &0.8246  \\
%KNN&0.7839 &0.7831 &0.7824  \\ 
%GaussianNaiveBayes&0.6956 &0.8876 &0.7789  \\ \hline
%\end{tabular}
%\label{tab:performance3}
%\end{table}

Furthermore, we validated the robustness of the \emph{Classifier} module with respect to the length of time windows. Intuitively, a larger time window could provide more information that could improve the detection accuracy, while a shorter time window might conserve time, which is beneficial for emergency response efforts. Therefore, striking a balance between the effectiveness and the efficiency of our framework constitutes an important discussion topic. Accordingly, we repeated the above cross-validation procedure for the \emph{Classifier} module with different post-window lengths, i.e., 5, 10, 15 and 20 seconds. The performances are shown in Figure~\ref{fig:timewindows}(a). Evidently, the performance of the \emph{Classifier} module is relatively stable with different time window lengths, indicating that our framework is not sensitive to the length of the cut time window.

\begin{table}[!t]
\center
\caption{\textbf{The performance of the \emph{EL-Picker} framework with continuous seismic waveforms.} Here, we demonstrate a performance comparison between the complete framework of \emph{EL-Picker} (i.e., \emph{Trigger} + \emph{Classifier} + \emph{Refiner}) and the partial framework in which the \emph{Classifier} module is hidden, when the length of post-window is set as 20s. ``CVn'' indicates the n-th fold of the cross-validation procedure, ``CV-Napa'' indicates the vlidation on Napa data set,
}
\setlength{\tabcolsep}{4.5mm}{
\begin{tabular}{ccccccc}
\hline\hline
\multirow{2}{*}{} & \multicolumn{3}{c}{\emph{Trigger} + \emph{Refiner}} & \multicolumn{3}{c}{\emph{EL-Picker}} \\ \cline{2-7}
&Precision&Recall &F-Score &Precision&Recall	 & F-Score\\\hline
CV1 &0.0017&0.9005 &0.0034 &0.3165&0.7926	& 0.4524\\
CV2 &0.0019&0.9102 &0.0039 &0.4163&0.7435	& 0.5337\\
CV3 &0.0018&0.9266 &0.0036 &0.4349&0.7317	& 0.5456\\
CV4 &0.0017&0.9221 &0.0034 &0.4167&0.7338	& 0.5316\\
\hline
CV-Napa &0.0013 &0.9147 &0.0026 &0.4800 &0.4412 &0.4598 \\
\hline\hline
%CV1(15s)&0.0017&0.9005 &0.0034 &0.3075&0.7885	& 0.4425 \\
%CV1(10s)&0.0017&0.9005 &0.0034 &0.3098&0.7807	& 0.4436\\
%CV1(5s)&0.0017&0.9005 &0.0034 &0.3067&0.7836	& 0.4408\\\hline
\end{tabular}}
\label{tab:performance2}
\end{table}

Next, we evaluated the overall performance of the complete \emph{EL-Picker} framework. Our experiments were conducted on complete and continuous seismic waveforms to simulate a real-world application scenario. In particular, if the time difference between one estimated P-phase arrival and one expert-labelled arrival was shorter than 0.4 seconds, we treated the estimated arrival as ``correct''. In this way, we repeated the above two cross-validation experiments within the complete framework of \emph{EL-Picker} (i.e., \emph{Trigger} + \emph{Classifier} + \emph{Refiner}) and with a baseline that was modified from our framework by hiding the \emph{Classifier} module (i.e.,\emph{ Trigger} + \emph{Refiner}). Indeed, the baseline could be regarded as a generalized version of the most widely used traditional approaches for identifying seismic P-phase arrivals. From the results summarized in Table~\ref{tab:performance2}, we find that \emph{EL-Picker} consistently outperforms the baseline with a significant margin (i.e., approximately 200 times) in terms of Precision, although the \emph{Classifier} module loses some performance in terms of Recall. Indeed, by comprehensively considering Precision and Recall (i.e., by considering F-score), the effectiveness of the \emph{EL-Picker} framework has been thoroughly validated. In addition, we also validate the robustness of our framework with various length of time windows for the \emph{Classifier} module, the results showed in Figure~\ref{fig:timewindows}(b) verify that our framework is not sensitive to the length of the cut time window.

In particular, compared with Table~\ref{tab:performance1}, the performance shown in Table~\ref{tab:performance2} seems worse. Moreover, several arrivals were captured by the \emph{EL-Picker} framework with a high confidence level but were not labelled by the experts as aftershocks. This phenomenon may have two potential explanations. First, compared with a simple classification task in which samples are picked in advance within a given time window, in the validations on continuous seismic waveforms, sometimes \emph{EL-Picker} captured one potential arrival, but it was beyond the pre-defined 0.4 second threshold; in this case, the real-time monitoring of continuously seismic waveforms could be much more difficult. Second, we neglected to consider that the manually labelled data set could contain some trivial omissions, leading to misjudgement. This issue will be discussed later in this article.

\begin{figure}[!b]
\centering
\begin{minipage}[c]{0.48\columnwidth}
\centering
\includegraphics[width=1.0\columnwidth ]{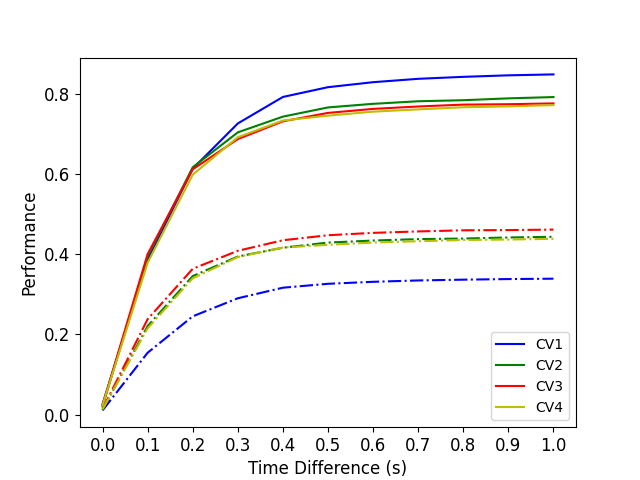}
\centerline{(a)}
\label{fig:CV-delta-1}
\end{minipage}
\hspace{0.02\textwidth}
\begin{minipage}[c]{0.48\columnwidth}
\centering
\includegraphics[width=1.0\columnwidth ]{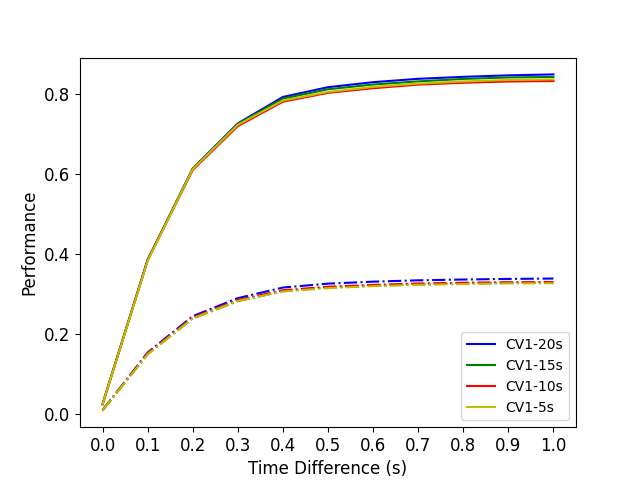}
\centerline{(b)}
\label{fig:CV-delta-2}
\end{minipage}
\caption{\textbf{The sensitivity to the time difference within the \emph{EL-Picker} framework. }  \textbf{(a)} The Precision (solid) and Recall (dashed) curves for the validations of \emph{EL-Picker} with different time difference boundaries. \textbf{(b)} The Precision (solid) and Recall (dashed) curves for the validations of \emph{EL-Picker} with different time difference boundaries and different post-window lengths in the ``CV1'' data set.}
\label{fig:CV-delta-time}
\vspace{-2mm}
\end{figure}

Next, we attempted to validate the sensitivity of \emph{EL-Picker} to the time difference. To that end, we evaluated different time difference boundaries to discriminate among the correct arrivals in the interval of [0 s, 1 s]. The performances of \emph{EL-Picker} with different settings are shown in Figure~\ref{fig:CV-delta-time}. Both Precision and Recall increase with increasing boundaries and tend to converge after 0.4 seconds, which means that the time differences between the identified P-phase arrivals and the expert-labelled arrivals are almost less than 0.4 seconds.

Finally, we turned to discuss the efficiency for our overall framework. To be specific, we selected the first week data set collected by one station as the testing data and run our framework on a single core Intel Xeon E5 2.2 GHz central processing unit to analyze the time consumption. Table~\ref{tab:timeconsumption} shows the number of signals each module processed and corresponding average running time for each signal.
We can find that the \emph{Trigger} module, based one STA/LTA model and equipped with low complexity, is capable to filter out most of noise with tiny running time. As a result, the \emph{Trigger} module enables our framework to avoid waste of computing resources, where only less than 0.08\% signal points are preserved and further processed by the \emph{Classifier} module, which has high computation complexity. 
Furthemore, the \emph{Classifier} module can also remove out 97.5\% false potential P-phase arrivals to reduce the unnecessary computation in the \emph{Refiner} module.
In addition, the modular manner of the \emph{Classifier} module and its parallelizable nature can also contribute to the time efficiency. For instance, the different categories of feature under various bandpass filters are independent and can be extracted in parallel manner. Similarly, the confidence score predication by each base model can be conducted simultaneously to guarantee the efficiency. Meanwhile, some base models are also developed based on the ensemble learning strategy and can be speeded up with multiple calculations, such as RandomForest and Adaboost.
Along this line, the running time of our \emph{Classifier} module relies on the maximal time consumption among the extraction processes for different feature categories and predication processes of different base models, not the sum of them. Along with this simple parallelization strategy, the average running time can be decreased into 17.2\% of that without any parallelization (more detailed discussions can be found in~\ref{SI:sec:Parallelization}).

\begin{table}[!t]
\center
\caption{\textbf{The time consumption of each module in the \emph{EL-Picker} framework.}
}
\setlength{\tabcolsep}{9mm}{
\begin{tabular}{cccc}
\hline\hline
                                                                     & Trigger  & Classifier & Refiner \\\hline
Number of Signals                                                        & 60480000 & 479204     &  12033       \\
\begin{tabular}[c]{@{}c@{}}Average Running Time / ms\end{tabular} & 0.0025   & 706.3258    &16.8616 \\
\hline \hline
\end{tabular}}
\label{tab:timeconsumption}
\end{table}

Based on the above results, both of the \emph{effectiveness} and the \emph{efficiency} of the \emph{EL-Picker} framework have been validated.

%\begin{figure}[!t]
%	\centering
%	\minipage[]{\label{fig:CV-delta-1}
%	\includegraphics[width=0.45\columnwidth ]{figs/delta-time-1}}
%	\minipage[]{\label{fig:CV-delta-2}
%	\includegraphics[width=0.45\columnwidth ]{figs/delta-time-2}}
%\caption{\textbf{The sensitivity to the time difference within the \emph{EL-Picker} framework. }  \textbf{(a)} The precision (solid) and recall (dashed) curves for the validations of \emph{EL-Picker} with different time difference boundaries. \textbf{(b)} Precision (solid) and recall (dashed) curves for the validations of \emph{EL-Picker} with different time difference boundaries and different post-window lengths in the ``CV1'' data set.}
%\label{fig:CV-delta-time}
%\end{figure}

%% file: Discussion.tex
\begin{figure}[!t]
\centering
\vspace{-0mm}
\includegraphics[width=1.0\columnwidth]{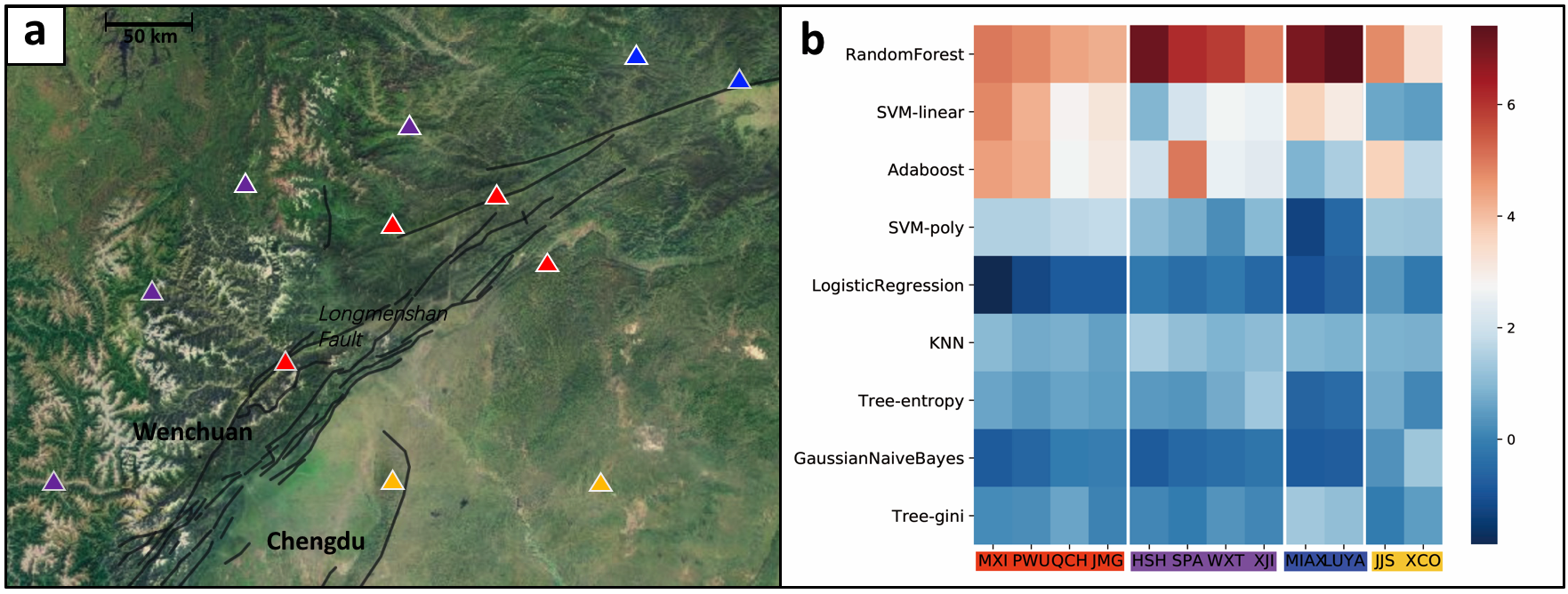}
\vspace{-0mm}
\caption{\textbf{The clustering results of 12 monitoring stations.}
 \textbf{(a)} The geographical distribution of 12 monitoring stations (triangles) around the Longmenshan Fault, which are divided into 4 clusters (distinguished by 4 different colours) according to their estimated parameters.  Obviously, the clustering results are strongly correlated with the geographical environments.  \textbf{(b)} A heat map diagram of the estimated parameters learned from the 12 stations, where each column represents a station and each row represents a base-model in the \emph{Classifier} module.}
%\vspace{-4mm}
\label{fig:clusterofstation} % \label should be placed after \caption
\end{figure}

\begin{figure}[!t]
\centering
\vspace{-0mm}
\includegraphics[width=1.0\columnwidth]{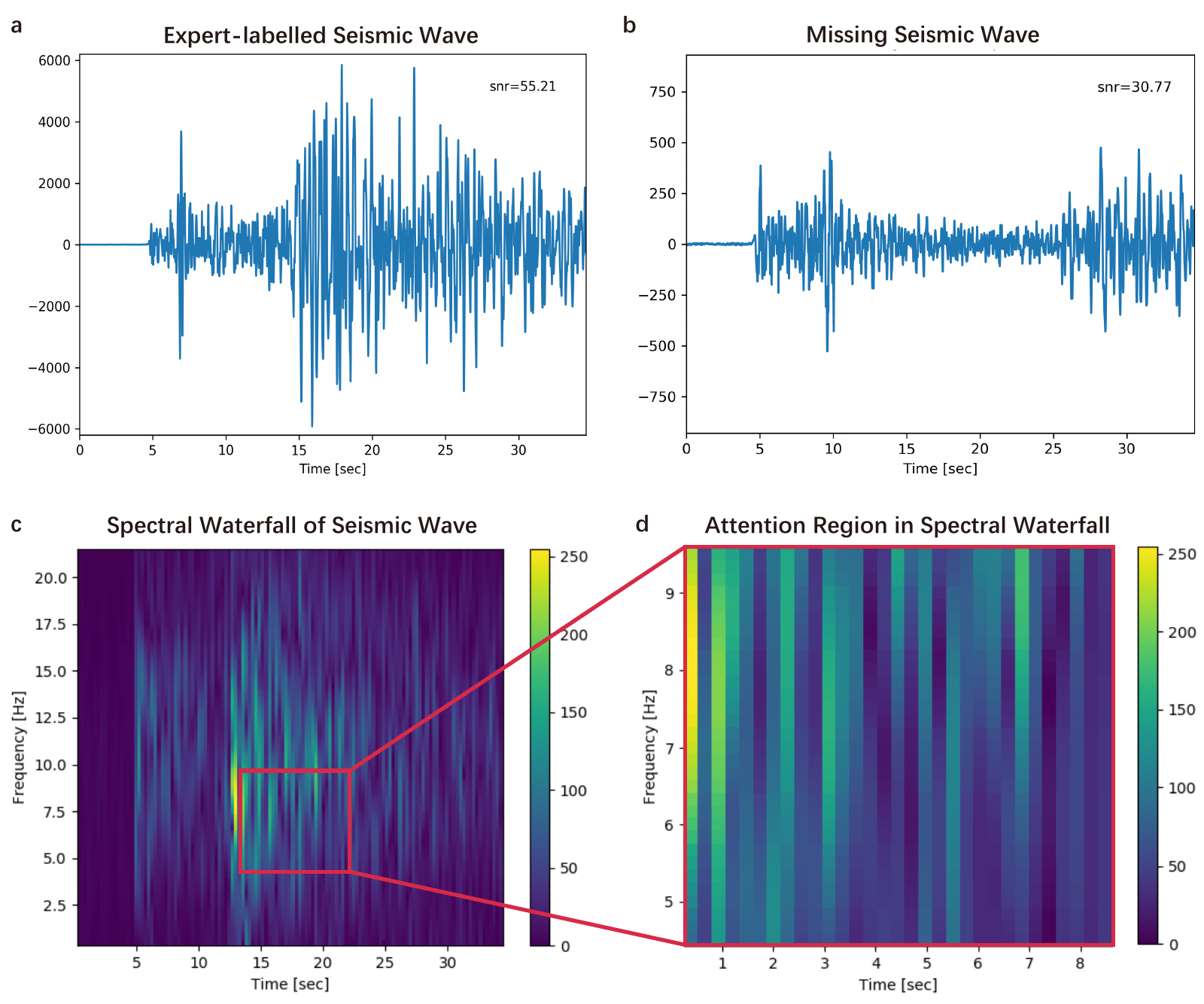}
\caption{\textbf{Case study on the statistics/model-based distinction between labelled and missing P-phases.}  The figures in the first row illustrate the average amplitude of each time window around the P-phase arrival for both labelled (left) and missing (right) P-phases. The figures in the second row illustrate the attention mask covering the spectral waterfall of an expert-labelled seismic P-phase arrival estimated by the RA-CNN and a magnified view of the peaks of S-phases, which should be investigated during this procedure.}
\vspace{-0mm}
\label{fig:differenceofmissing} % \label should be placed after \caption
\end{figure}
\section{Discussion}\label{sec:discussion}
As we have validated the effectiveness and efficiency of \emph{EL-Picker}, we now verify additional characteristics of the \emph{EL-Picker} framework to further ensure its applicability in real-world scenarios. Indeed, due to the differences among various geographical environments, the seismic waveforms recorded by different stations may contain different characteristics, even for the same earthquake. Therefore, we would like to investigate whether the \emph{EL-Picker} framework can be reasonably applied for various types of seismic waveforms. The \emph{Classifier} module of \emph{EL-Picker} is intuitively designed in a modular manner with ensemble learning to guarantee the scalability of identification compared with individual machine learning models. To that end, we designed a case study on the Wenchuan data set. More specifically, For each monitoring station in the Wenchuan data set, we separately trained the model with the data from all four weeks to estimate the personalized parameters of each station as a 9-dimensional vector (the detailed personalized parameters for each station are shown in Table~\ref{tab:stationCluster} in~\ref{SI:sec:detailinformation}). Afterwards, based on their personalized parameters, we divided all the 12 stations into 4 clusters by the K-means~\cite{hartigan1979algorithm} method, which is effective for clustering analysis. 
From the results shown in Figure~\ref{fig:clusterofstation}, we find that the red cluster is located along the centre of the Longmenshan Fault, while the purple cluster is surrounded by mountains, the blue cluster lies along the edge of the mountain range and faults, and the yellow cluster is far from the faults and lies within the plain. Obviously, the clustering results as well as the personalized parameter settings are highly correlated with the geographical environment, which correspond to the different models selected in \emph{EL-Picker}. For example, for the stations near the faults (i.e., the red cluster), the weights of RandomForest~\cite{ho1995random}, SVM-linear and Adaboost\cite{freund1997decision} are significantly higher than those of the other base models. For the stations throughout the mountains (i.e., the purple cluster and the blue cluster), RandomForest is the most prominent choice, while for the stations on the plain (i.e., the yellow cluster), the weights of the base models vary in a smaller range.

Next, we will discuss a special issue observed in the previous validations. As mentioned above, for each experiment
in the second validation, several arrivals were captured by the \emph{EL-Picker} framework with a high confidence level but were not labelled by the experts as aftershocks. To ensure the applicability of \emph{EL-Picker}, we have to check whether the technical framework needs further refinement or if the manually labelled data set contains some trivial omissions. Correspondingly, we randomly selected 1,000 samples (arrivals) with a higher confidence than the threshold from the seismic sequences. Then, an expert from the CEA was asked to review these samples with exceptional scrutiny and caution to reduce the false alarm rate. According to the results, 907 captured P-phases among the 1,000 random samples were judged as aftershocks, further validating the effectiveness of \emph{EL-Picker}. However, 125.6\% more P-phases  than the expert-labelled data were ignored during the first round of labelling, which raises the new challenge for explaining the ignorance during manual inspection.

We first counted the average amplitude for each arrival and its time window, including a 5-second pre-window and a 30-second post-window. As shown in the first row of Figure~\ref{fig:differenceofmissing}, the differences between the expert-labelled earthquakes and missing aftershocks have the potential to be dramatically significant. More specifically, among the expert-labelled earthquakes (the left figure in the first row), the waveforms exhibit violent fluctuations at approximately 15-25 seconds after the earthquakes. In contrast, at the corresponding periods in the missing aftershocks (the right figure in the first row), the waveforms are relatively stable with a lower signal-to-noise ratio (SNR). In these cases, the secondary phase arrivals, i.e., after 25 seconds, could be even more distinct, which suggests a long distance between the epicentre and monitoring stations (more examples are shown in Figure~\ref{fig:exampleforautoandlabel} in~\ref{SI:sec:detailinformation}). 
A similar rule could also be revealed by the modelling process; we utilized the recurrent attention (RA)-CNN model~\cite{fu2017look} to train a classifier to differentiate the spectral waterfalls of expert-labelled and missing P-phase arrivals to discover their significant differences. 
The RA-CNN is a neural network-based classification algorithm and can find an attention region in the spectral waterfall to obtain better classification results than can be achieved with the full-size spectral waterfall. (more technical details are shown in~\ref{SI:sec:racnn}) As shown in the second row of Figure~~\ref{fig:differenceofmissing}, in which the attention mask covers the spectral waterfall, the waveforms near the peaks of S-phases, i.e., at approximately 15-25 seconds, should be especially investigated to discover missing aftershocks.

According to these results, we conclude that regional earthquakes with insignificant waveforms around the peaks of S-phases can be easily ignored during manual labelling, especially when large earthquakes occur; furthermore, manual labelling can be an urgent yet laborious task for geophysical experts. Consequently, significant earthquakes may be identified first with the highest priority, while insignificant earthquakes may be ignored. This phenomenon is certainly reasonable: we always tend to finish the easiest questions in a quiz and leave the hardest questions for the end or even abandon them due to time limitations. At the same time, we also verify that \emph{EL-Picker} could effectively assist experts in retrieving missing earthquakes, thereby diminishing the heavy burden of the labelling task.

In summary, with two layers of modular capabilities, i.e., the assembly of the \emph{Trigger}, \emph{Classifier} and \emph{Refiner} modules and the design of the \emph{Classifier} module with ensemble learning consisting of multiple machine learning techniques, the \emph{EL-Picker} framework has been verified as a competitive solution for seismic monitoring applications due to its effectiveness and efficiency, as well as its flexibility and stability. Moreover, since \emph{EL-Picker} can identify low-SNR earthquakes, the monitoring performance can be further refined with reduced manual burden.

%% file: Conclusion.tex
\section{Conclusion}
In this paper, we introduced a machine learning-enhanced framework, \emph{EL-Picker}, for the automatic identification of seismic P-phase arrivals on continuous and massive waveforms. A unique perspective of \emph{EL-Picker} is that it consists of three modules, namely, \emph{Trigger}, \emph{Classifier}, and \emph{Refiner}, and an ensemble learning strategy is exploited to integrate several machine learning classifiers. Extensive evaluation of the aftershocks following the $MS$ 8.0 Wenchuan earthquake demonstrated that \emph{EL-Picker} can not only achieve the best identification performance but also identify 120\% more seismic P-phase arrivals as complementary data. Meanwhile, experiments also revealed both the applicability of different machine learning models for waveforms collected from different seismic stations and the regularities of seismic P-phase arrivals that might be neglected during manual inspection. These findings clearly validated the \emph{effectiveness}, \emph{efficiency}, \emph{flexibility} and \emph{stability} of \emph{EL-Picker}.

%% file: supportinformation.tex
\section*{Supplementary Information}
\setcounter{figure}{0}
\makeatletter
\renewcommand{\thefigure}{S\@arabic\c@figure}
\renewcommand{\thetable}{S\@arabic\c@table}
\makeatother
\subsection*{Technical Details for RA-CNN.}\label{SI:sec:racnn}
In Discussion section, to re-inspect the P-phase arrivals, we utilized the recurrent attention (RA)-CNN model[50] to train a classifier to differentiate the spectral waterfalls of expert-labelled and missing P-phase arrivals to discover their significant differences. Compared with the original network, our RA-CNN consists of two stages. The inputs of the first stage are the full-size spectral waterfalls of the seismic waveforms, whereas the inputs of the second stage are the attention regions. More specifically, in different stages, spectral waterfalls were fed into three convolutional layers with depths of 16, 32, and 64; in the first stage, a region-based feature representation was extracted by an additional attention proposal network. Then, the results of both stages were evaluated to predict the probability scores by using one 128-unit fully connected layer and a softmax layer. The proposed RA-CNN was optimized for convergence by alternatively learning the softmax classification loss in each stage and a pairwise ranking loss across neighbouring stages. We used $L2$ regularization with normalization on each layer and used the Adam optimizer to train the network.

\subsection*{Detail information for the Parallelization of the Classifier Module.}\label{SI:sec:Parallelization}
Here, we display more details about the time consumption analysis for our \emph{Classifier} module. 
As Table~\ref{tab:time_feature} shows, different feature categories with different bandpass filters can be extracted in parallel, and the running time can be decreased into the maximal time cost among them. Similarly, as Table~\ref{tab:time_basemodel} shows, in the predication process, the overall running time with parallelization should amount to the sum of the maximum of time cost for all base models and the time cost for  meta model. Meanwhile, the RandomForest and Adaboost are also developed based on ensemble learning strategy, where the time cost of predication can be decreased by a factor of the number of parallel processes. Therefore, the overall running time of the Classifier module in parallel mainly relies on the time cost of KNN model. As a result, the average running time for the \emph{Classifier} module in parallel is 121.7502  ms, 17.2\% of that without any parallelization.

\begin{table}[!htpb]
\center
\caption{\textbf{The time consumption of feature extraction in the \emph{Classifier} module}}
\setlength{\tabcolsep}{9mm}{
\begin{tabular}{ccc}
\hline  \hline
Feature Extraction    & Running Time / ms            & Number of Filters \\ \hline
Amplitude Fluctuation & 26.5335                 & 2                  \\
Maximal Amplitude     & 2.7649                 & 2                  \\
Spectral Waterfall    & 45.5803                & 9                  \\
Other Features        & 97.7476                & 5                  \\ \hline\hline
                      & Without Parallelization & In Parallel        \\ \hline
Overall Running Time / ms & 172.6262              & 19.5495         \\ \hline \hline
\end{tabular}}
\label{tab:time_feature}
\end{table}

\begin{table}[!htpb]
\center
\caption{\textbf{The time consumption of each base model and meta model in the \emph{Classifier} module}}
\setlength{\tabcolsep}{6mm}{
\begin{tabular}{cccc}
\hline \hline
&model            & Running Time / ms            & Ensemble Learning  \\ \hline
\multirow{9}{*}{base model}&SVM-linear            & 5.1925                & FALSE              \\
&SVM-ploy              & 5.9858                & FALSE              \\
&NaiveBayes            & 0.4301               & FALSE              \\
&KNN                   & 102.2007                & FALSE              \\
&LogisticRegression    & 0.1689               & FALSE              \\
&RandomForest          & 303.9377                & TRUE               \\
&AdaBoost              & 115.5387                & TRUE               \\
&Tree-gini             & 0.0785               & FALSE              \\
&Tree-entropy          & 0.0734               & FALSE              \\ \hline
\multirow{1}{*}{meta model}&LogisticRegression    & 0.0934              & FALSE              \\ \hline \hline
\multicolumn{2}{c}{}              & Without Parallelization & In Parallel        \\ \hline
\multicolumn{2}{c}{Overall Running Time / ms}  & 533.6996               & 102.2007         \\ \hline \hline
\end{tabular}}
\label{tab:time_basemodel}
\end{table}

\subsection*{More Detail Information Mentioned in our Article}\label{SI:sec:detailinformation}
\begin{figure}[!h]
	\centering
\begin{minipage}[c]{0.45\columnwidth}
	\includegraphics[width=1.0\columnwidth ]{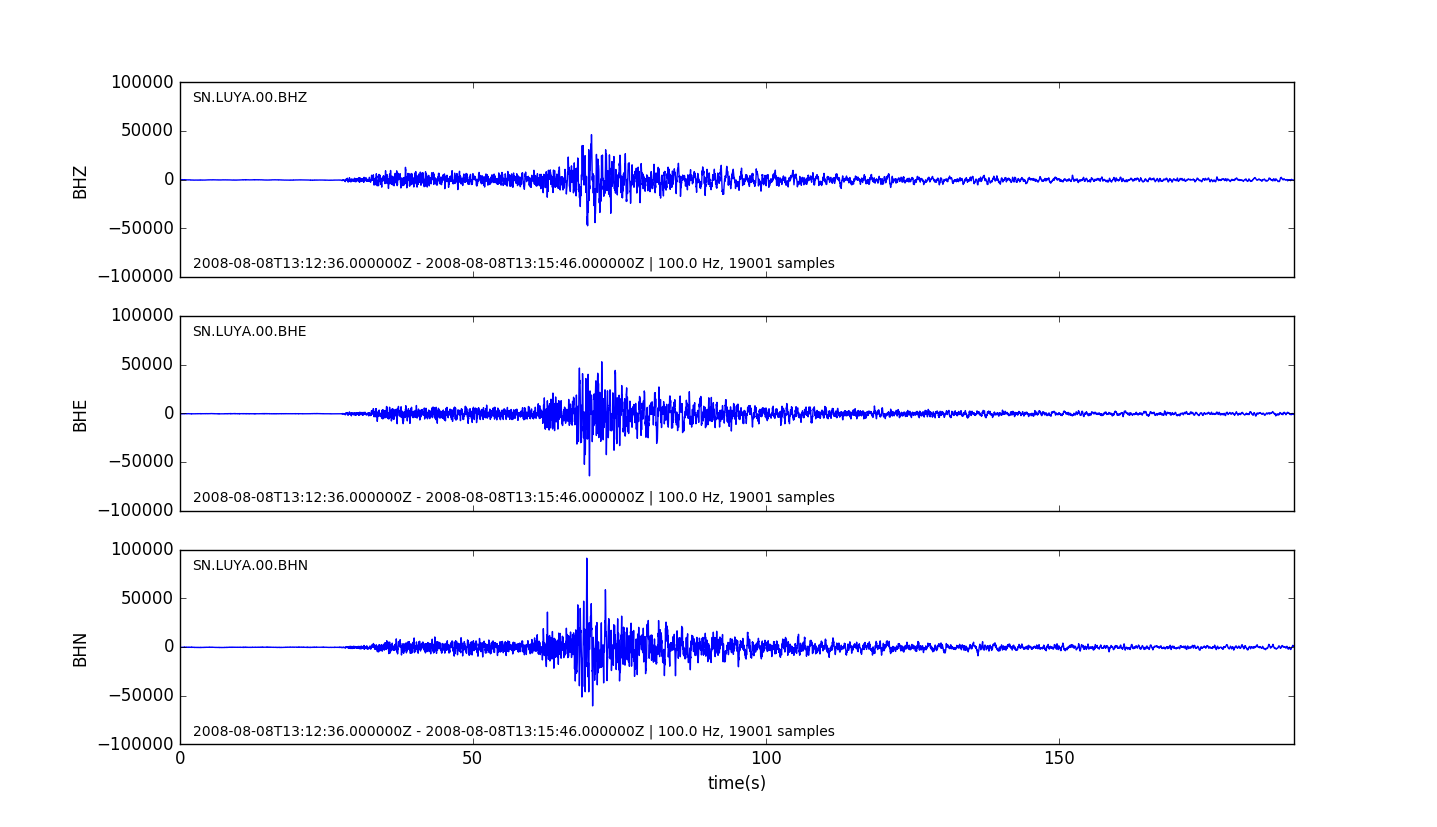}
	\centerline{(a)}
\end{minipage}
\begin{minipage}[c]{0.45\columnwidth}
	\includegraphics[width=1.0\columnwidth ]{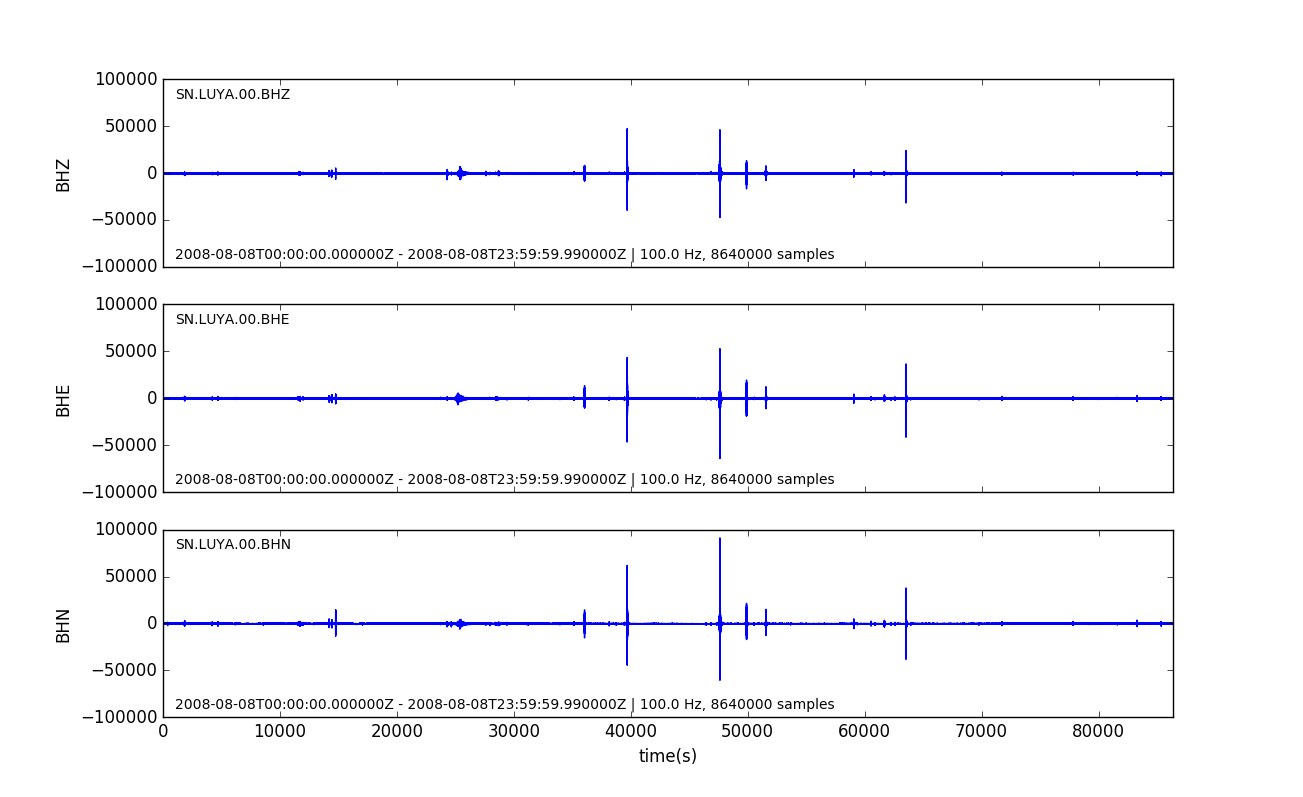}
	\centerline{(b)}
\end{minipage}
\caption{\textbf{Examples for waveform data in three channels.}\textbf{(a)} A waveform example of a seismic phase. \textbf{(b)} A waveform example for one whole day.}
\label{fig:examplefordata}
\end{figure}

\begin{figure}[!tphb]
	\centering
	\includegraphics[width=0.45\columnwidth ]{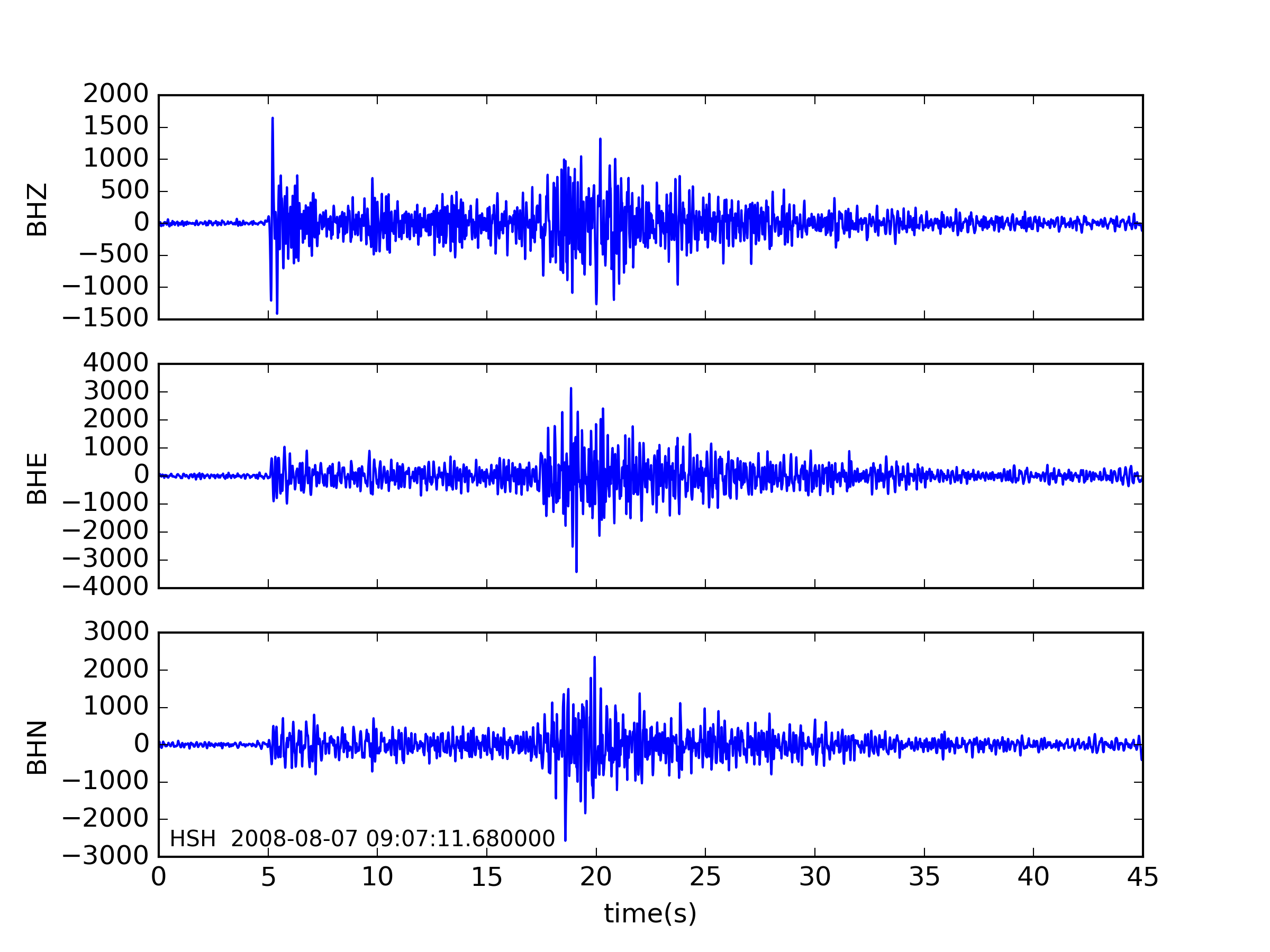}
	\includegraphics[width=0.45\columnwidth ]{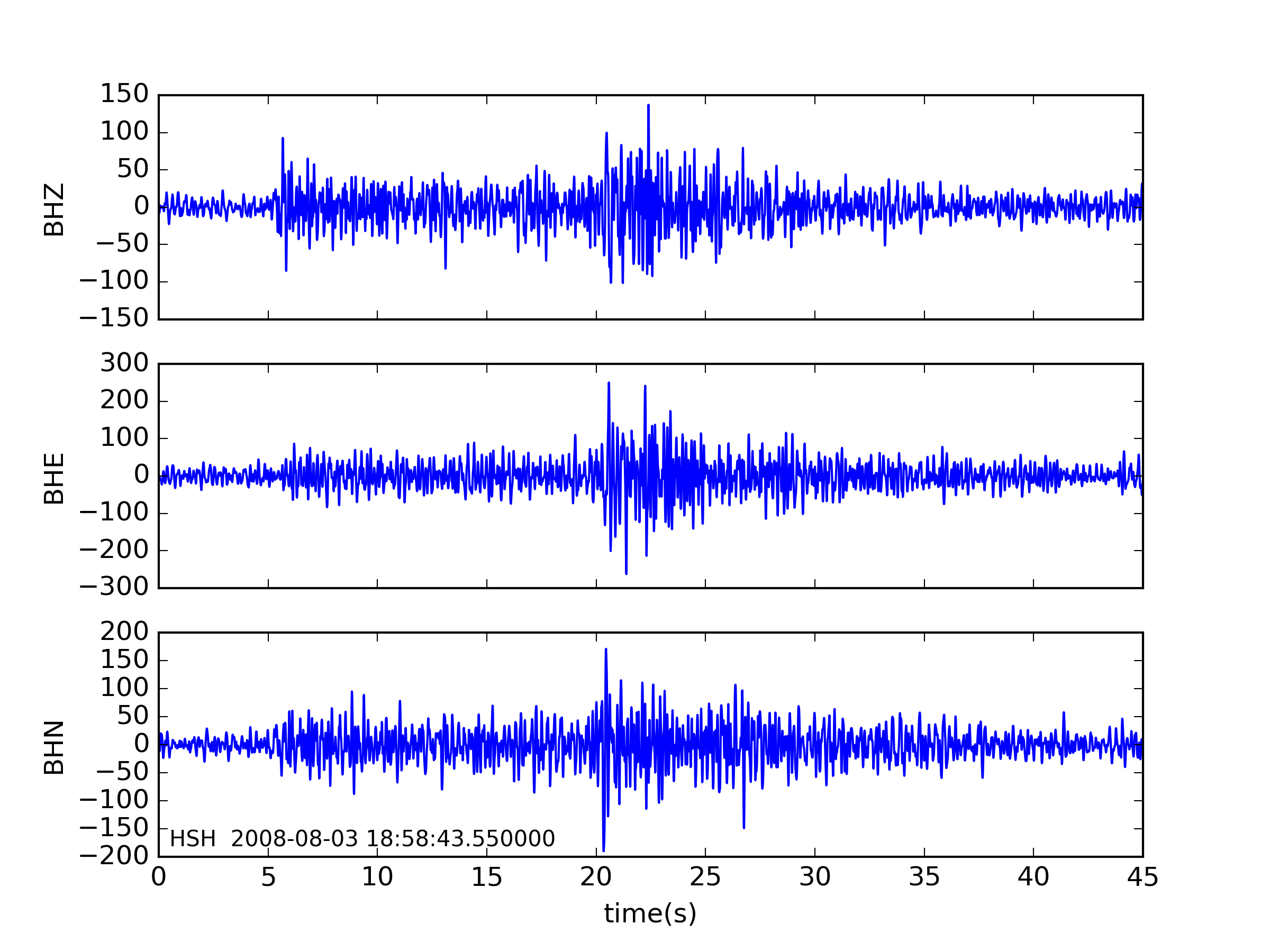}
\begin{minipage}[c]{0.45\columnwidth}
	\includegraphics[width=1.0\columnwidth ]{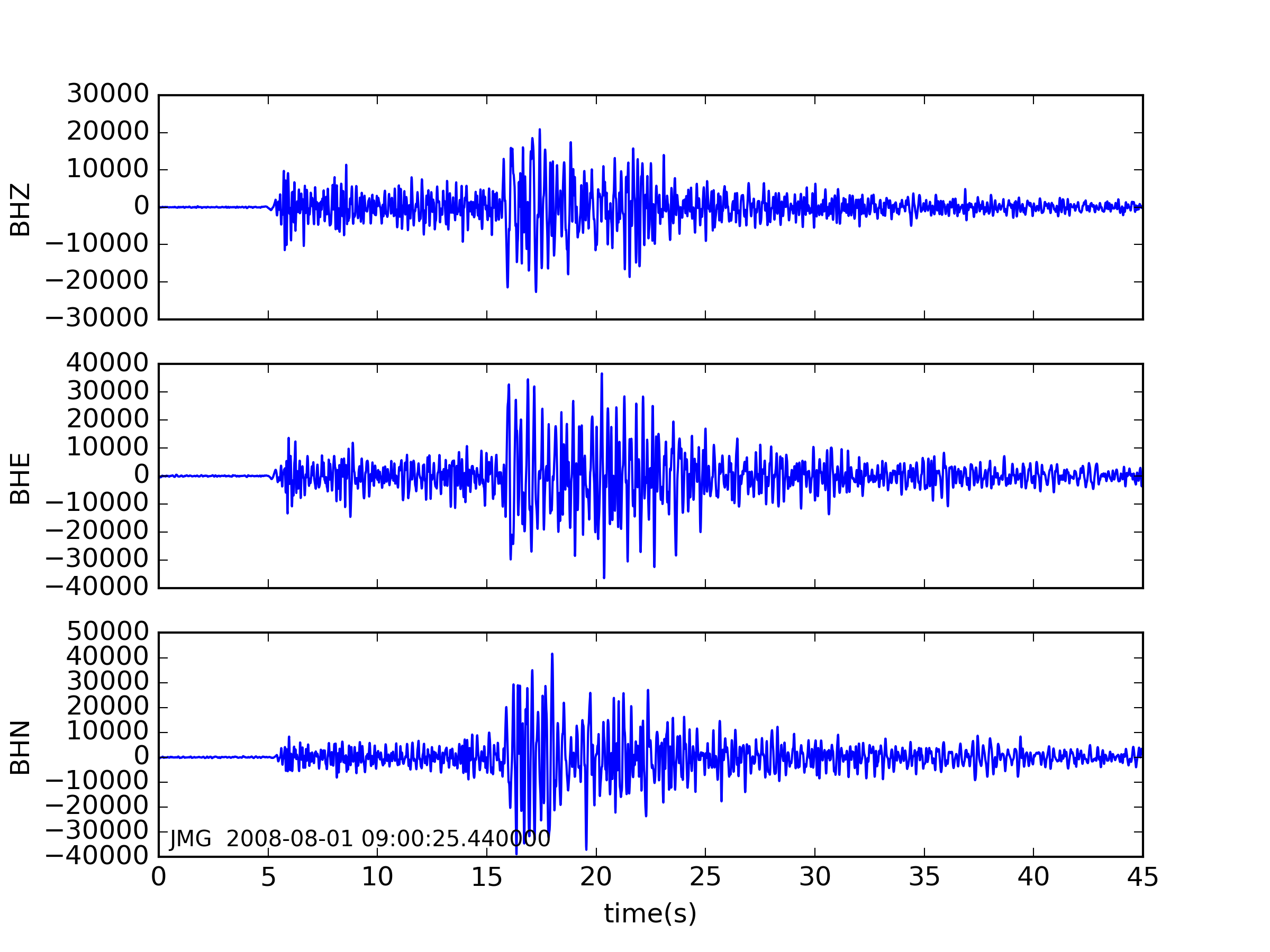}
	\centerline{(a) Expert-labeled P-phases}
\end{minipage}
%\hspace{0.02\textwidth}
\begin{minipage}[c]{0.45\columnwidth}
	\includegraphics[width=1.0\columnwidth ]{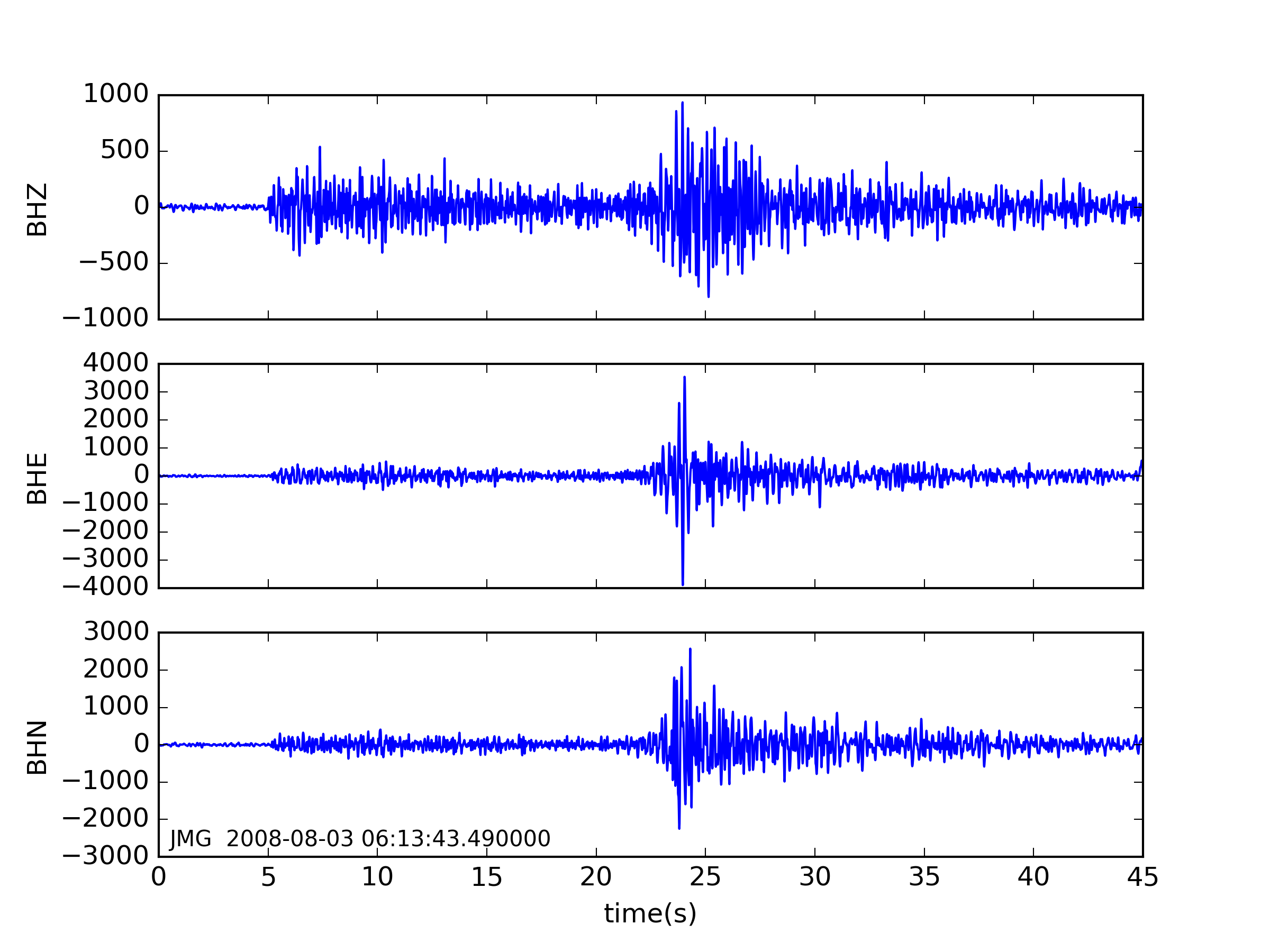}
	\centerline{(b) Missing P-phases automatic picked by \emph{EL-Picker}}
\end{minipage}
	\caption{\textbf{Examples for expert-labeled P-phases and missing P-phases.}\textbf{(a)} Waveform examples around expert-labeled P-phase arrivals. \textbf{(b)} Waveform examples around missing P-phase arrivals automatic picked by \emph{EL-Picker}. All P-phase arrivals are located at 5 seconds.}
\label{fig:exampleforautoandlabel}
\end{figure}

In this section, we introduce some detail information mentioned in our article. To be specific, the data used in this paper consists of aftershocks of the $MS$ 8.0 Wenchuan earthquake in the time period from 01/08/2008 to 28/08/2008,and aftershocks of the $MW$ 6.0 Napa earthquake in the time period from 24/08/2014 to 27/08/2014. Meanwhile, in our experiments these 28 days related to Wenchuan dataset were divided into 4 weeks: 01/08-07/08, 08/08-14/08, 15/08-21/08 and 22/08-28/08. The waveform data set was collected from 12 seismic stations around Wenchuan, namely MXI, PWU, QCH, JMG, MIAX, LUYA, HSH, SPA, WXT, XJI, JJS and XCO. Supplementary figures will show some examples for seismic waveform data (Figure~\ref{fig:examplefordata}) mentioned in section~\ref{sec:framework} and expert-labeled P-phases and missing P-phases (Figure~\ref{fig:exampleforautoandlabel}) mentioned in section~\ref{sec:discussion}. Supplementary Tables contain details for feature functions (Table~\ref{tab:functionsforfeature}) mentioned in section~\ref{sec:data}, and the personalized parameters for each monitoring station in 4 clusters mentioned in section~\ref{sec:discussion} (Table~\ref{tab:stationCluster}).
\newpage
\begin{table}[!h]
\centering
  %\fontsize{4.9}{6}\selectfont
  \caption{\textbf{Detail for Features Functions.} Here, functions $mean(x)$, $max(x)$ $var(x)$ aim to compute the average, maximum and variance of the amplitudes (absolute value) in time window $x$.  Function $sum(y)$ sums each element in vector $y$. Function $x.index(a)$ aims to find the position of value $a$ in time window $x$. Function $cov(v)$ outputs the covariance matrix of vector $v$. Function $eign(mv)$ tries to compute the all eigenvalues of matrix $mv$.}
	\renewcommand\arraystretch{1.5}    
    \resizebox{0.75\linewidth}{!}{
    \begin{tabular}{|c|l|}
    \hline
    Function & Detail for functions\\ \hline \hline
    Normalization(x) & \makecell[l]{\textbf{Input: time window x}\\ \textbf{Output: $\frac{x - mean(x)}{sqrt(var(x))}$}}\\ \hline
    Root Mean Square(RMS) Amplitude Ratio(x,y) & \makecell[l]{\textbf{input: two time windows $x,y$}\\ \textbf{Output: $\frac{sum([t^2~for~t~in~x])}{sum([t^2~for~t~in~y])}$}}\\ \hline
    Mean Difference(x, y) & \makecell[l]{\textbf{Input: two time windows $x,y$} \\ \textbf{Output $mean(x)-mean(y)$}}\\ \hline
    Envelope Slope(x) & \makecell[l]{\textbf{Input: time window $x$ with range (-5, 5)} \\a = $max(x[-5,-1.5])$\\
    $ta = x[-5,-1.5].index(a)$\\
    $b = max(x[1.5:5])$\\
    $tb = x[1.5:5].index(b)$\\
    $c = max(x[-0.5:0.5])$\\
    $tc = x[-0.5:0.5].index(c)$\\
    \textbf{Output: $\frac{c - a}{(tc - ta)}$, $\frac{c - b}{100(tc - tb)}$}}\\ \hline
    Polarization Slope(a,b,c)& \makecell[l]{\textbf{Input:the amplitudes of the arrival in} \\ \textbf{three channels, $a,b,c$ } \\ $v=vector(a,b,c)$ \\ $mv=covariance(v)$ \\ $a,b,c=eigen(mv)$\\ \textbf{Output: $\frac{(a-b)^2+(a-c)^2+(b-c)^2}{2(a+b+c)^2}$}} \\ \hline
    \end{tabular}}
\label{tab:functionsforfeature}
\end{table}

\newpage
\begin{table}[!t]
  \centering
  %\fontsize{4.9}{6}\selectfont
  \normalsize
  \renewcommand\arraystretch{1.5}
  \caption{\textbf{The personalized parameters for each monitoring station in 4 clusters.} For each station cluster, the corresponding table shows the personalized parameters for each station and the average of absolute values of those parameters, which imply the weight for each base-model selected during the ensemble processing. And, the three most preferred base-models are highlighted for each cluster.}
    %\begin{tabular}{|p{4cm}<{\centering}|p{1.5cm}<{\centering}|p{2cm}<{\centering}|}
   \resizebox{1.0\linewidth}{!}{
   \begin{tabular}{|c|c|c|c|c|c|}
    \hline
\makecell{Stations in Cluster 1}&MXI&PWU&QCH&JMG& \makecell{Average of \\absolute value} \\ \hline \hline
\textbf{RandomForest}&4.9942 &4.7718 &4.3815 &4.2266 &\textbf{4.5935}  \\ \hline
\textbf{SVM-linear}&4.7747 &4.1946 &2.8170 &3.1989 &\textbf{3.7463}  \\ \hline
\textbf{Adaboost}&4.4733 &4.2846 &2.6682 &3.0870 &\textbf{3.6283}  \\ \hline
SVM-poly&1.5487 &1.5619 &1.7213 &1.7963 &1.6570  \\ \hline
LogisticRegression&-1.8847 &-1.2002 &-0.7953 &-0.8197 &1.1750  \\ \hline
KNN&1.0137 &0.7089 &0.8086 &0.5722 &0.7759  \\ \hline
Tree-entropy&0.6527 &0.4312 &0.5975 &0.4989 &0.5451  \\ \hline
GaussianNaiveBayes&-0.7675 &-0.5447 &-0.0783 &-0.0039 &0.3486  \\ \hline
Tree-gini&0.1982 &0.2500 &0.6519 &0.0479 &0.2870  \\ \hline
    \end{tabular}
    \begin{tabular}{|c|c|c|c|}
    \hline
 \makecell{Stations in Cluster 2} &MIAX&LUYA&ll\makecell{Average of \\ absolute value} \\ \hline \hline
\textbf{RandomForest}&6.9713 &7.4213 &\textbf{7.1963} \\ \hline
\textbf{SVM-linear}&3.6626 &3.0471 &\textbf{3.3548} \\ \hline
\textbf{Tree-gini}&1.3582 &1.1351 &\textbf{1.2467} \\ \hline
Adaboost&0.9006 &1.4879 &1.1942 \\ \hline
KNN&0.9600 &0.8543 &0.9072 \\ \hline
SVM-poly&-1.2885 &-0.5111 &0.8998 \\ \hline
LogisticRegression&-0.9975 &-0.6844 &0.8409 \\ \hline
GaussianNaiveBayes&-0.8234 &-0.7510 &0.7872 \\ \hline
Tree-entropy&-0.6483 &-0.4533 &0.5508 \\ \hline
    \end{tabular}}
\label{tab:stationCluster}
\end{table}

\begin{table}[!h]
  \centering
  %\fontsize{4.9}{6}\selectfont
  \normalsize
  \renewcommand\arraystretch{1.5}
  %\caption{\textbf{The personalized parameters for each monitoring station in 4 clusters.} }
    %\begin{tabular}{|p{2cm}|p{1.5cm}|p{1.5cm}|p{1.5cm}|p{1cm}|}
  \resizebox{1.0\linewidth}{!}{
    \begin{tabular}{|c|c|c|c|c|c|}
    \hline
\makecell{Stations in Cluster 3}&HSH&SPA&WXT&XJI& \makecell{Average of \\absolute value} \\ \hline \hline % \\ \large{\color[rgb]{0.433,0.207,0.617}$\blacktriangle$}
\textbf{RandomForest}&7.1496 &6.1390 &5.8661 &4.8817 &\textbf{6.0091}  \\ \hline
\textbf{Adaboost}&2.0263 &5.0003 &2.5758 &2.3877 &\textbf{2.9975}  \\ \hline
\textbf{SVM-linear}&0.9171 &2.1159 &2.6915 &2.5828 &\textbf{2.0768} \\ \hline
KNN&1.4380 &1.1525 &0.9095 &1.0815 &1.1454  \\ \hline
SVM-poly&1.0604 &0.8022 &0.2247 &1.0578 &0.7863  \\ \hline
Tree-entropy&0.4588 &0.3758 &0.7509 &1.3359 &0.7303  \\ \hline
GaussianNaiveBayes&-0.7671 &-0.5272 &-0.3661 &-0.2216 &0.4705  \\ \hline
LogisticRegression&-0.1256 &-0.3749 &-0.1349 &-0.5175 &0.2882  \\ \hline
Tree-gini&0.1292 &-0.0976 &0.3664 &0.1313 &0.1811  \\ \hline
    \end{tabular}
    \begin{tabular}{|c|c|c|c|}
    \hline
\makecell{Stations in Cluster 4 } &JJS&XCO& \makecell{Average of \\absolute value} \\ \hline \hline %\\ \large{\color[rgb]{0.918,0.699,0.234}$\blacktriangle$}
\textbf{RandomForest}&4.7374 &3.3000 &\textbf{4.0187}  \\ \hline
\textbf{Adaboost}&3.6855 &1.7180 &\textbf{2.7017}  \\ \hline
\textbf{SVM-poly}&1.3044 &1.2725 &\textbf{1.2884} \\ \hline
KNN&0.8719 &0.8371 &0.8545  \\ \hline
GaussianNaiveBayes&0.3171 &1.3104 &0.8138  \\ \hline
SVM-linear&0.6842 &0.5038 &0.5940  \\ \hline
Tree-entropy&0.7271 &0.1163 &0.4217  \\ \hline
Tree-gini&-0.0676 &0.5094 &0.2885  \\ \hline
LogisticRegression&0.4118 &-0.1383 &0.2750  \\ \hline
    \end{tabular}}
%\label{tab:stationCluster}
\end{table}

\newpage

%% file: main.bbl
\begin{thebibliography}{99}

\begin{CJK}{UTF8}{gbsn}
\bibitem{shen2020machinelearning} Shen D, Zhang Q, Xu T. A machine learning-enhanced robust P-phase picker for real-time seismic monitoring,  SCIENTIA SINICA Informationis, 2020 [申大忠，张琦，徐童等， EL-Picker：基于集成学习的余震P波初动实时拾取方法，中国科学：信息科学， 2020]， DOI: \href{http://engine.scichina.com/doi/10.1360/SSI-2020-0214}{10.1360/SSI-2020-0214}. 
\end{CJK}
\bibitem{fang2017seismOlympics} Fang L, Wu Z, Song K. Seismolympics, Seismological Research Letters, 2017, 88: 1430-1430

\bibitem{gibbons2006detection} Gibbons S J, Ringdal F, The detection of low magnitude seismic events using array-based waveform correlation, Geophysical Journal International, 2006, 165, 1: 149-166

\bibitem{shelly2007non} Shelly D R, Beroza G C, Ide S. Non-volcanic tremor and low-frequency earthquake swarms, Nature, 2007, 446: 7133

\bibitem{plenkers2013low} Plenkers K, Ritter J R, Schindler M, Low signal-to-noise event detection based on waveform stacking and cross-correlation: application to a stimulation experiment, Journal of seismology, 2013, 17: 27-49

\bibitem{saragiotis1999higher} Saragiotis C D, Hadjileontiadis L J, Panas S M. A higher-order statistics-based phase identification of three-component seismograms in a redundant wavelet transform domain, In: Proceedings of the IEEE Signal Processing Workshop on Higher-Order Statistics, 1999,  396-399

\bibitem{saragiotis2002pai} Saragiotis C D, Hadjileontiadis L J, Panas S M. Pai-s/k: A robust automatic seismic p phase arrival identification scheme, IEEE Transactions on Geoscience and Remote Sensing, 2002, 40: 1395-1404

\bibitem{kuperkoch2010automated} Ku L, Meier T, Lee J, et al. Automated determination of p-phase arrival times at regional and local distances using higher order statistics, Geophysical Journal International, 2010, 181: 1159-1170

\bibitem{allen1978automatic} Allen R. Automatic earthquake recognition and timing from single traces, Bulletin of the Seismological Society of America, 1978, 68: 1521-1532

\bibitem{allen1982automatic} Allen R. Automatic phase pickers: Their present use and future prospects, Bulletin of the Seismological Society of America, 1982, 72: S225-S242

\bibitem{lomax2012automatic} Lomax A, Satriano C, Vassallo M. Automatic picker developments and optimization: Filterpicker—a robust, broadband picker for real-time seismic monitoring and earthquake early warning, Seismological Research Letters, 2012, 83: 531-540

\bibitem{ruano2014seismic} Ruano A E, Madureira G, Barros O, et al. Seismic detection using support vector machines, Neurocomputing, 2014, 135: 273-283

\bibitem{beyreuther2008continuous} Beyreuther M, Wassermann J. Continuous earthquake detection and classification using discrete hidden markov models, Geophysical Journal International, 2008, 175: 1055-1066

\bibitem{wang1995artificial} Wang J, Teng T L. Artificial neural network-based seismic detector, Bulletin of the Seismological Society of America, 1995, 85: 308-319

\bibitem{dai1997application} Dai H, MacBeth C. The application of back-propagation neural network to automatic picking seismic arrivals from single-component recordings, Journal of Geophysical Research: Solid Earth, 1997, 102: 15105-15113

\bibitem{ross2018generalized} Ross Z E, Meier M A, Hauksson E, et al. Generalized seismic phase detection with deep learning, arXiv preprint arXiv:180501075, 2018

\bibitem{ross2018p} Ross Z E, Meier M A, Hauksson E. P-wave arrival picking and first-motion polarity determination with deep learning, Journal of Geophysical Research: Solid Earth, 2018

\bibitem{perol2018convolutional} Perol T, Gharbi M, Denolle M. Convolutional neural network for earthquake detection and location, Science Advances, 2018, 4: 2

\bibitem{zhu2018phasenet} Zhu W, Beroza G C. Phasenet: a deep-neural-network-based seismic arrival-time picking method, Geophysical Journal International, 2018, 216: 261-273

\bibitem{zhang2019aftershock} Zhang Q, Xu T, Zhu H, et al. Aftershock Detection with Multi-scale Description Based Neural Network. In: Proceedings of 2019 IEEE International Conference on Data Mining. IEEE, 2019: 886-895.

\bibitem{mousavi2019cred} Mousavi S M, Zhu W, Sheng Y, et al. CRED: A deep residual network of convolutional and recurrent units for earthquake signal detection. Scientific Reports, 2019, 9(1): 1-14.

\bibitem{wiszniowski2014application} Wiszniowski J, Plesiewicz B M, Trojanowski J. Application of real time recurrent neural network for detection of small natural earthquakes in Poland. Acta Geophysica, 2014, 62(3): 469-485.

\bibitem{chen2016xgboost} Chen T, Guestrin C. Xgboost: A scalable tree boosting system. In: Proceedings of the 22nd ACM SIGKDD International Conference on Knowledge Discovery and Data Mining. 2016: 785-794.

\bibitem{brown2008autocorrelation} Brown J R, Beroza G C, Shelly D R. An autocorrelation method to detect low frequency earthquakes within tremor, Geophysical Research Letters, 2008, 35:16

\bibitem{aguiar2014pagerank} Aguiar A C, Beroza G C. Pagerank for earthquakes, Seismological Research Letters, 2014, 85: 344-350

\bibitem{yoon2015earthquake} Yoon C E, O’Reilly O, Bergen K J, et al. Earthquake detection through computationally efficient similarity search, Science advances, 2015, 1: 11

\bibitem{baer1987automatic} Baer M, Kradolfer U. An automatic phase picker for local and teleseismic events. Bulletin of the Seismological Society of America, 1987, 77(4): 1437-1445.

\bibitem{earle1994characterization} Earle P S, Shearer P M. Characterization of global seismograms using an automatic-picking algorithm. Bulletin of the Seismological Society of America, 1994, 84(2): 366-376.

\bibitem{wu2019research} Wu Y, Wei J, Pan J, et al. Research on Microseismic Source Locations Based on Deep Reinforcement Learning. IEEE Access, 2019, 7: 39962-39973.

\bibitem{asim2017earthquake} Asim K M, Martínez-Álvarez F, Basit A, et al. Earthquake magnitude prediction in Hindukush region using machine learning techniques. Natural Hazards, 2017, 85(1): 471-486.

\bibitem{riggelsen2014machine} Riggelsen C, Ohrnberger M. A machine learning approach for improving the detection capabilities at 3c seismic stations, Pure and Applied Geophysics, 2014, 171: 395-411

\bibitem{zhu2020rapid} Zhu H, Sun Y, Zhao W, et al. Rapid Learning of earthquake felt Area and intensity Distribution with Real-time Search engine Queries. Scientific Reports, 2020, 10(1): 1-9.

\bibitem{gentili2006automatic} Gentili S, Michelini A. Automatic picking of P and S phases using a neural tree. Journal of Seismology, 2006, 10(1): 39-63.

\bibitem{graves2013hybrid} Graves A, Jaitly N, Mohamed A. Hybrid speech recognition with deep bidirectional LSTM. In: Proceedings of 2013 IEEE Workshop on Automatic Speech Recognition and Understanding. IEEE, 2013: 273-278.

\bibitem{rokach2010ensemble} Rokach L. Ensemble-based classifiers, Artificial Intelligence Review, 2010, 33: 1-39

\bibitem{wolpert1992stacked} Wolpert D H. Stacked generalization, Neural networks, 1992, 5: 241-259

\bibitem{cristianini2000introduction} Cristianini N, Shawe-Taylor J, et al., An introduction to support vector machines and other kernel-based learning methods, Cambridge university press, 2000

\bibitem{cox1958regression} Cox D R. The regression analysis of binary sequences, Journal of the Royal Statistical Society Series B (Methodological), 1958, 215-242

\bibitem{breiman2017classification} Breiman L. Classification and regression trees, Routledge, 2017

\bibitem{ho1995random} Ho T K. Random decision forests, In: Proceedings of the third international conference on document analysis and recognition, 1995. 278-282

\bibitem{freund1997decision} Freund Y, Schapire R E. A decision-theoretic generalization of on-line learning and an application to boosting, Journal of Computer and System Sciences, 1997, 55: 119-139

\bibitem{russell2016artificial} Russell S J, Norvig P. Artificial intelligence: a modern approach, Malaysia; Pearson Education Limited,2016

\bibitem{kohavi1995study} Kohavi R, et al. A study of cross-validation and bootstrap for accuracy estimation and model selection, In: Precossing of International Joint Conferences on Artificial Intelligence Organization, Montreal, Canada, 1995. 1137-1145

\bibitem{akaike1974new} Akaike H. A new look at the statistical model identification, IEEE transactions on automatic control, 1974, 19: 716-723

\bibitem{sleeman1999robust} Sleeman R, Van Eck T. Robust automatic p-phase picking: an on-line implementation in the analysis of broadband seismogram recordings, Physics of the earth and planetary interiors, 1999, 113: 265-275

\bibitem{leonard1999multi} Leonard M, Kennett B. Multi-component autoregressive techniques for the analysis of seismograms, Physics of the Earth and Planetary Interiors, 1999, 113: 247-263

\bibitem{leonard2000comparison} Leonard M. Comparison of manual and automatic onset time picking, Bulletin of the Seismological Society of America, 2000, 90: 1384-1390

\bibitem{hochreiter1997long} Hochreiter S, Schmidhuber J. Long short-term memory. Neural Computation, 1997, 9(8): 1735-1780.

\bibitem{xie2017aggregated} Xie S, Girshick R, Dollár P, et al. Aggregated residual transformations for deep neural networks. In: Proceedings of the IEEE Conference on Computer Vision and Pattern Recognition. 2017: 1492-1500.

\bibitem{hartigan1979algorithm} Hartigan J A, Wong M A. A k-means clustering algorithm, Journal of the Royal Statistical Society Series C (Applied Statistics), 1979, 28: 100-108

\bibitem{fu2017look} Fu J, Zheng H, Mei T. Look closer to see better: Recurrent attention convolutional neural network for fine-grained image recognition, In: Processing of Conference on Computer Vision and Pattern Recognition, 2017. 3

\end{thebibliography}
